\documentclass[conference]{IEEEtran}
\usepackage[numbers, square]{natbib}
\usepackage{amsmath,amssymb,amsfonts,todonotes}
\usepackage{algorithmic,comment}
\usepackage{graphicx}
\usepackage{url}

\usepackage{breakurl}
\makeatletter % changes the catcode of @ to 11
\newcommand{\linebreakand}{%
  \end{@IEEEauthorhalign}
  \hfill\mbox{}\par
  \mbox{}\hfill\begin{@IEEEauthorhalign}
}
\makeatother % changes the catcode of @ back to 12
\usepackage{tikz}

\usepackage{caption}
\usepackage{subcaption}

\usepackage{pgfplots}

\usepackage[linesnumbered,ruled,vlined]{algorithm2e}

\usepackage{textcomp}
\usepackage{xcolor}

\usepackage{multirow}

\usepackage{amsthm}

\usepackage{adjustbox}
\usepackage[nointegrals]{wasysym}
\usepackage{booktabs}
\usepackage{pdfcomment}
\usepackage{siunitx}
\usepackage{float}

\def\BibTeX{{\rm B\kern-.05em{\sc i\kern-.025em b}\kern-.08em
    T\kern-.1667em\lower.7ex\hbox{E}\kern-.125emX}}
    
\begin{document}

\SetKwInput{KwInput}{Input}                % Set the Input
\SetKwInput{KwOutput}{Output}              % set the Output

%Conference

%Commands for commenting:
%\newcommand{\says}[3]{\todo[size=\small,color=#2,inline]{#1 says: #3}}
\newcommand{\says}[3]{}

\newcommand\luis[1]{\says{Luis}{red}{#1}}
\newcommand\eric[1]{\says{Eric}{cyan}{#1}}
\newcommand\yi[1]{\says{Yi}{green}{#1}}
\newcommand{\nils}[1]{\says{Nils}{lightgray}{#1}}
\newcommand{\saman}[1]{\says{Saman}{red}{#1}}
\newcommand{\matt}[1]{\says{Matt}{magenta}{#1}}

\newcommand{\orange}[1]{\textcolor{orange}{#1}}
\newcommand{\green}[1]{\textcolor{green}{#1}}
\newcommand{\blue}[1]{\textcolor{blue}{#1}}
\newcommand{\red}[1]{\textcolor{red}{#1}}

% Commands for title
\newcommand{\name}{\textsc{EvilEye\hspace{3pt}}}
\newcommand{\namenospace}{\textsc{EvilEye}}
\newcommand{\nametitle}{EvilEye}

% Commands for table
%https://tex.stackexchange.com/questions/32683/rotated-column-titles-in-tabular
\newcommand*\rot{\multicolumn{1}{R{60}{1em}}}
\newcommand*\rota{\multicolumn{1}{R{90}{1em}}}
\newcommand{\Tdot}{$\CIRCLE$}
\newcommand{\Thdot}{$\LEFTcircle$}
\newcommand{\Twdot}{$\Circle$}
\newcommand{\etal}{\emph{et al.}\ }
\newcolumntype{R}[2]{%
	>{\adjustbox{angle=#1,lap=\width-(#2)}\bgroup}%
	l%
	<{\egroup}%
}

%\title{Deception is in the Eye of the Beholder: \\Optical Man-in-the-Middle Physical Attacks} 
%\title{\nametitle: Man-in-the-middle Perception Attacks with Adversarial Optics}
%Why don't you clean your glasses? Perception attacks via dynamic optical perturbations
\title{Why Don't You Clean Your Glasses? Perception Attacks with Dynamic Optical Perturbations}
%\title{Why Don't You Clean Your Glasses?\\Perception Attacks with Dynamic Optical Perturbations}
% \title{Blinded by the Lights\\Perception Attacks with Dynamic Optical Perturbations}

% Manufactured Mirages: (Augmenting Perception | Augmented Perception Attacks) with Adversarial Optics

% Double Vision: ...

% bring in constrained additive physical pertubations
% evading object detection through constrained adversarial physical pertubations
% Fata Morgana

%\eric{would it make sense to include the phrase "optical man-in-the-middle attacks" somewhere in the title? Maybe "Dynamic Physical Attacks on Camera-based Autonomous Systems via Optical Man-in-the-Middle Attacks"??}}
%\author{Anonymous Author(s)}

\author{\IEEEauthorblockN{Yi Han}
\IEEEauthorblockA{\textit{Rutgers University} \\
\textit{yh482@rutgers.edu}}
\and
\IEEEauthorblockN{Matthew Chan}
\IEEEauthorblockA{\textit{Rutgers University} \\
\textit{matthew.chan@rutgers.edu}}
\and
\IEEEauthorblockN{Eric Wengrowski}
\IEEEauthorblockA{\textit{Steg AI} \\
\textit{eric.wengrowski@gmail.com}}
\and
\IEEEauthorblockN{Zhuohuan Li}
\IEEEauthorblockA{\textit{Rutgers University} \\
\textit{zl253@scarletmail.rutgers.edu}}
\and
\IEEEauthorblockN{Nils Ole Tippenhauer}
\IEEEauthorblockA{\textit{CISPA Helmholtz Center for Information Security} \\
\textit{tippenhauer@cispa.de}}
\and
\IEEEauthorblockN{Mani Srivastava}
\IEEEauthorblockA{\textit{University of California, Los Angeles} \\
\textit{mbs@ucla.edu}}
\and
\IEEEauthorblockN{Saman Zonouz}
\IEEEauthorblockA{\textit{Georgia Institute of Technology} \\
\textit{saman.zonouz@gatech.edu}}
\and
\linebreakand
\IEEEauthorblockN{Luis Garcia}
\IEEEauthorblockA{\textit{University of Utah} \\
\textit{lgarcia@cs.utah.edu}}

}
\newcommand{\figurepath}{Figures}

% \and
% \and
% \and
% \and
% \and
\maketitle
\begin{abstract}
Camera-based autonomous systems that emulate human perception are increasingly being integrated into safety-critical platforms.
Consequently, an established body of literature has emerged that explores adversarial attacks targeting the underlying machine learning models. 
% These attacks produce high-confidence misclassifications by digitally manipulating pixels in an image, taking advantage of the complex and non-convex functions learned by these classifiers.
%These attacks take advantage of incongruities in the learned manifolds of classifiers by carefully manipulating image pixels to produce misclassifications with high confidence, and are usually generated and applied digitally. %\eric{``incongruities in the learned manifolds of classifiers'' is accurate but reads slightly weird, like we're trying to sound fancy} 
Adapting adversarial attacks to the physical world is desirable for the attacker, as this removes the need to compromise digital systems. However, the real world poses challenges related to the ``survivability" of adversarial manipulations given environmental noise in perception pipelines and the dynamicity of autonomous systems.
%Modeling these attacks in the real, physical world is desired because they eliminate compromised camera requirements, but they are much more challenging given the environmental noise in perception pipelines and the dynamicity of autonomous systems. 
%\eric{we don't explicitly mention the wide-range of existing security practices to protect against digital-domain attacks, minimizing the attack surface}\luis{It may be too much for the abstract. It's already long as is. We also shouldn't have a paragraph break.}
%However, these digital domain attacks require access to electronic systems that may be rigorously secured. 
%While effective, one practical drawback of these attacks is the system access required to digitally manipulate classifier inputs.
%Adversaries have recently taken advantage of static, physical-domain attacks focusing on camera-observed objects. However, these static attacks on physical objects have limited practicality and may be susceptible to human-in-the-loop detection.
%To circumvent this, recent work focuses on attacks in the physical domain involving the malicious modification of camera-observed objects. These physical adversarial attacks don't require direct access to classifier inputs, but are usually static with limited portability, and may be susceptible to human-in-the-loop detection.
%\eric{reason for a paragraph break here?}
In this paper, we take a sensor-first approach.
%We demonstrate that a small device placed on a camera sensor can be used to perform dynamic adversarial attacks and evade human-in-the-loop detection.
%We demonstrate how dynamic attacks can exploit the optics of cameras to produce targeted misclassification for several objects under a variety of illumination conditions.
We present \namenospace, a man-in-the-middle perception attack that leverages transparent displays to generate dynamic physical adversarial examples. \name exploits the camera's optics to induce misclassifications under a variety of illumination conditions.
%Not only can we generate dynamic payloads, but we also formalize the transformation of a digital attack projected onto the physical domain by modeling the optics of the camera-scene transformation function.
To generate dynamic perturbations, we formalize the projection of a digital attack into the physical domain by modeling the transformation function of the captured image through the optical pipeline. %\eric{are ``dynamic payloads'' and ``camera scene'' the best way to describe this?}
%%We focus on an autonomous driving use case, where camera sensors detect and classify nearby objects to inform vehicle decisions. 
%Our extensive experiments show that \name generates robust adversarial perturbations that can mislead both classifiers and object detectors with up to $70\%$ attack success rate (ASR).  
% \eric{$99\%$ success is a very bold claim. how true is this figure?}
Our extensive experiments show that \namenospace's generated adversarial perturbations are much more robust across varying environmental light conditions relative to existing physical perturbation frameworks, achieving a high attack success rate (ASR) while bypassing state-of-the-art physical adversarial detection frameworks. %\eric{is ``light conditions'' clear in this context?}\luis{Reworded.} %This makes the attack a bigger threat in a real world scenario. 
We demonstrate that the dynamic nature of \name enables attackers to adapt adversarial examples across a variety of objects with a significantly higher ASR compared to state-of-the-art physical world attack frameworks. 
% that the dynamic nature of \name enables attackers to adapt adversarial examples across a va
%$30\%$ higher ASR compared to state-of-the-art, static approaches. 
%\eric{couple of things: 1. replace ``all kinds of traffic signs'' with something more precise, like ``a set of 10 roadways signs as defined by the  Manual on Uniform Traffic Control Devices prepared by the US Department of Transportation Federal Highway Administration Office of Transportation Operations'' (or cite this manual). 2. this is the first time that we mention the self-driving car application, so instantly jumping into the fact that we can attack more traffic signs than some uncited state-of-the-art approach is confusing without first laying out the proper context. 3. ``enables it to attack'' sounds weird. does \name perform the attack, or does an attacker utilize \name? we should make a decision and be consistent throughout the paper.}\luis{Attempted a fix} 
%\name is able to bypass the state-of-the-art physical adversarial detection framework. %\eric{no citation in the abstract?}\luis{The abstract should be self-contained without citations. I removed the Sentinet name.} 
Finally, we discuss mitigation strategies against the \name attack. % how the \name framework can be leveraged at training time to make existing defenses more robust. 
%\eric{this is confusion for 2 reasons. 1, it's a new item that follows a sentence already qualified with ``finally.'' 2. if this is a defense, then we should call it a defense rather than dancing around it.}\luis{It should read better now. I don't think we should explicitly label our approach as a defense since a large majority of our evaluation focuses on attacks.}

%We also find our generated attacks can evade detection by SentiNet, a framework for identifying physical adversarial examples.

% Our code and data have been made publicly available.
\end{abstract}

% From SLAP:
% We study the feasibility of SLAP in the self-driving scenario, targeting both object detector and traffic sign recognition tasks, focusing on the detection of stop signs. We conduct experiments in a variety of ambient light conditions, including outdoors, showing how in non-bright settings the proposed method generates AE that are extremely robust, causing misclassifications on state-of-the-art neural networks with up to 99% success rate. Our experiments show that SLAP-generated AE do not present detectable behaviours seen in adversarial patches and therefore bypass SentiNet, a physical AE detection method. We evaluate other defences including an adaptive defender using adversarial learning which is able to thwart the attack effectiveness up to 80% even in favourable attacker conditions.

%\begin{IEEEkeywords}
%cyber-physical systems, adversarial machine learning
%\end{IEEEkeywords}

\section{Introduction}\label{sec:intro}
% Prevalence of Deep learning and subsequent attacks

%First paragraph:Intro to DL
Safety-critical autonomous \yi{would the word autonomous gives the reviewers the impression that this is about autonomous vehicles?} systems commonly rely on optical sensors coupled with artificial intelligence to mimic human perception across a variety of domains, including autonomous navigation~\cite{pendleton2017perception}, crowd surveillance and analysis~\cite{sreenu2019intelligent}, and facial recognition for authentication~\cite{mehdipour2016comprehensive}. %Humans can interpret images with ease; however replicating this capability on autonomous systems using artificial intelligence algorithms is a rigorous challenge with a vast body of existing research~\cite{}. Recently, these algorithms have made their way into many industrial and consumer-facing applications. A prime example of this is the automotive industry, which has been increasingly adopting a camera-first paradigm \cite{} to replace more complex or expensive LIDAR and ultrasonic sensor approaches~\cite{takacs2018highly}. Deep learning especially has been established as a popular class of algorithms for computer vision tasks~\cite{}. The ability to simultaneously learn latent features and classifiers through optimization in the training process has made feed-forward deep learning an attractive option when sufficient training data exists.
%But there are risks associated with relying on artificial intelligence for safety-critical systems.
Although the associated deep learning algorithms enable promising performance across domains~\cite{goodfellow2016deep}, they introduce vulnerabilities that are non-obvious and can be targeted by adversaries. %\eric{1. why are the associated ML algorithms powerful? can we use a more precise term than ``powerful''? 2. citations needed for ``powerful algorithms'', ``difficult to understand vulnerabilities'', and ``can be targeted by adversaries''.}
There have already been real-world instances where perceptual models have failed, resulting in fatalities~\cite{banks2018driver}.
Researchers have shown various autonomous sensors to be exploitable~\cite{yan2016can}, even production vehicles such as Tesla cars can be compromised~\cite{deng2020analysis}. %\eric{1. ``a Tesla automobile'' vs. ``Tesla automobiles''. 2. anything can be targeted. should read ``successfully attacked'' or something similar}

Much of the initial research on adversarial machine learning~\cite{huang2011adversarial} has focused on digital-domain software attacks, a holdover from more thoroughly-explored computer vision research.
More recent work has explored the feasibility of adversarial machine learning attacks on sensors in the physical domain, where dynamic environmental conditions make precision attacks more difficult to execute~\cite{eykholt2018robust,chen2018shapeshifter,athalye2018synthesizing,thys2019fooling,zhao2019seeing,wu2020making}.
%\matt{todo: rewrite following paragraph to fit new table 1. Additionally, table 1 is described later on as well in S2B.}
%Existing physical-domain approaches can be broadly classified along two axes, as exemplified in Table~\ref{tab:relatedWorks}.
%On one axis, attacks can range in their proximity to the digital domain: either augmenting perceived objects themselves to attack autonomous vision systems, or augmenting the sensor pipeline just before an image is fed to the software. On the other axis, attacks can be static or dynamic in nature, being either fixed and context independent, or adaptive using contextual and environmental information.
%As shown in Table~\ref{tab:relatedWorks}, we categorize physical domain attacks based on the location of the perturbation, i.e., object-level modification (OM) versus sensor-level modification (SM), as well as the dynamicity of the attack, i.e., static attacks (SA) calculated ahead of time versus dynamic attacks (DA) that can change at runtime. Additionally, we distinguish online attacks (OA) that generate dynamic perturbations based on the current context of the attack. 

The development of robust, physical-domain attacks has trended from static, object-level attacks~\cite{eykholt2018robust,chen2018shapeshifter,athalye2018synthesizing,thys2019fooling,zhao2019seeing,wu2020making, jia2022fooling, man2020ghostimage, wang2021daedalus, wang2021can} to dynamic, object-level attacks~\cite{hoory2020dynamic,lovisotto2020slap,nichols2018projecting,nassi2020phantom} for misclassifying specific objects for non-static domains, e.g., autonomous vehicles detecting objects in various environments. However, object-level approaches hinge on augmented objects entering the target perception pipeline, e.g., an autonomous vehicle passing by a maliciously modified stop sign. Moreover, to attack multiple objects, each object needs to be modified individually. To overcome the aforementioned problems, recent attacks~\cite{ji2021poltergeist, kohler2021they, man2020ghostimage, wang2021can, li2019adversarial, zolfi2021translucent} have started exploring sensor-modifying attacks. These attacks typically exploit vulnerable sensors (e.g. cameras, CMOS sensors or inertial sensors) in the perception pipeline of the victim system. Most relevant to this paper, in~\cite{li2019adversarial, zolfi2021translucent} showed that stickers with static, adversarial perturbations could be placed on the lens of a target camera. However, such an approach cannot adapt to various scenes and environmental conditions due to the static nature of the stickers. %\luis{I modified the previous paragraph significantly. I tried to sumarize the limitations for each cell in Table~\ref{tab:attack-cats}}%Although existing attacks show how vulnerable autonomous vision systems can be with impressive results, these attacks have limitations. \eric{citation?} Attacks that augment specific target objects cannot easily scale to multiple target objects. \eric{will it be clear to the reader why?} Also, it is more difficult for adversarial perturbations applied on the objects to survive strong light conditions. \eric{this sentence comes way out of left field, and no citation! Context is needed for a bold claim like this!} Static attacks have the disadvantage that once a perturbation is generated and applied, it cannot be changed. \eric{citation??} The aforementioned limitations significantly reduce the threats of existing attack approaches in real-world scenarios.

\begin{comment}
    What is the risk associated with these attacks? The latter example resides in the domain of adversarial machine learning (AML) attacks. Initial research focused on the efficacy of AML attacks only in the digital domain~\cite{}, which may be secured through traditional means. Recent work has explored the feasibility of AML attacks in the physical domain, where dynamic and complex circumstances make precise attacks more difficult~\cite{}. The physical-domain approaches can be broadly categorized as \textit{static} or \textit{dynamic} attacks, and orthogonally attacks that target at an \textit{object-level} or attacks that target the \textit{autonomous sensor pipeline}. Table~\ref{tab:attack-cats} summarizes the different classes and corresponding examples of physical AML attacks. While \textit{dynamic} attacks may change based on the underlying system state, \textit{static} attacks do not. \textit{Object-level} attacks are defined as modifications to one or more specific physical objects in an environment. Attacks on \textit{autonomous sensor pipelines} target the systems of sensors and autonomous algorithms designed to perform and react to state estimations of their environment. 
\end{comment}
%\eric{These definitions are a bit wordy. Any suggestions on how to simplify?}

% In this paper, we aim to generate more intelligent model to attack the sensor pipeline
In this paper, we fill the gap by proposing an intelligent model, \namenospace, to dynamically attack the sensor pipeline. \name utilizes a portable transparent display in front of an optical sensor to display a carefully crafted adversarial perturbation to alter the perception of the sensor. 
% -- depicted in Figure~\ref{fig:attack-overview}
An auxiliary sensor is utilized to detect when a target object, e.g., a ``Stop" sign, enters the victim camera's field of view, which dynamically informs the generation of an adversarial perturbation. To craft adversarial perturbations, \name first models the optical sensor pipeline, i.e., the optical path of a digitally generated adversarial perturbation from the transparent display to the optical sensor, using a feed-forward neural network. 
%\eric{this is an important step worth emphasizing. why can't I just generate a digital perturbation without modeling the optical sensor pipeline? because our attack has to account for the spectral sensitivity of the camera sensor, the spectral emmitance of the transparent display, and the radiometric properties of light in free space~\cite{wengrowski2016optimal,wengrowski2017reading}. Rather than modeling each of these components individually, a camera-display transfer function (CDTF) jointly models these component systems. Following Wengrowski et al., a trained neural network models the CDTF~\cite{wengrowski2019light}.} 
The sensor pipeline model enables \name to simulate physical adversarial attacks by implicitly accounting for the spectral sensitivity of the camera sensor, the spectral emittance of the transparent display, and the radiometric properties of light in free space~\cite{wengrowski2016optimal,wengrowski2017reading}. \name then utilizes a gradient-based iterative approach to generate adversarial perturbations. To make sure adversarial perturbations can survive dynamic conditions in a physical environment, \name simulates various environmental factors (e.g., perspective, distance, and illuminance) when searching for an adversarial perturbation. 

\name has the following advantages over prior attack frameworks: 1) our attack does not require knowledge of when and where a target object will enter the periphery of a target sensor since perturbation decisions can be made at runtime, whereas prior works require a target sensor to follow a particular path; 
2) our attack is more flexible than prior approaches because the adversarial perturbations can be adjusted dynamically (e.g., targeting different traffic signs for misclassification). According to our extensive empirical evaluations, this results in better attack performance; and 3) our attack is robust across various environmental lighting conditions.
%\matt{double-check the following sentence}
%This is an important factor for any light-based attack, as additional ambient light negatively affects the relative attack strength.
%Due to our perturbation's close physical proximity to the optical sensor, it is less affected by environmental light. %\eric{need to first explain why environmental light conditions matters. could briefly describe the threat model from a sensor perspective, or cite state-of-the-art work where environmental lighting has a significant effect} \eric{is \name a model or an attack? currently using both in this paragraph.} 
%\matt{todo: no more face, right? (possible discussion point?)}
We empirically evaluate \name in the traffic sign recognition domain. %traffic sign classification and face verification. 
We will make our code open source for reproduction and future research.

% We aim to decompose the various components of the adversarial sample generation pipeline to model the physical sensor pipeline separately from the training data distribution.

% We further aim to show this in real-time
%Fourth Paragraph: real-time attacks
% Because the physical world is chaotic, we similarly show how to make this system robust to changes in the physical domain.
% \begin{itemize}
% \item{Here we need to distinguish ourselves from the real-time adversarial attack paper} 
% \item{We first start with static objects (targets) and move on to dynamic objects}
% \end{itemize}

\noindent\textbf{Contributions:} The contributions set forth in the paper are summarized as follows:
\begin{itemize}
  \item We introduce a new class of dynamic, sensor-first  machine learning attacks in the physical domain. %\eric{will readers know what ``AML'' means?}
  \item We formalize the contextualization of adversarial examples in the optical sensor pipeline. %\luis{The SLAP paper does this; need to position contrib. better. Something like, "Contextualization of adversarial examples in the optical sensor pipeline."}
  \item We develop and validate a novel, compact device designed to attack the optics of a camera system.
  \item We extensively evaluate the efficacy of our sensor-first attack on a traffic sign recognition task across various levels of illumination, significantly outperforming existing approaches.%for safety-critical autonomous driving, as well as facial recognition for authentication. % use-cases.%\luis{SLAP paper also does this. Something along of the lines of autonomous driving car driving through a city on a non-deterministic path with diverse objects/targets. More generally, evaluate on applications where adversary may have changing target objects.}
%   \item Introduction of the new \textbf{NAMED} dataset of $N$ images used to train our proposed attacks.
%   \item Public release of all code used to run the experiments presented in this paper. \luis{Latter two contributions are not contributions, but we can add them as sentences after.}
\end{itemize}

\begin{table}[t]
	\centering 
	\caption{A comparison of the characteristics of closely related works. \Tdot \  indicates taking this approach,\  \LEFTcircle \ indicates the work could plausibly take this approach, and \ \Twdot \ means that the work does not utilize the approach. OM: Object-Modifying; SM: Sensor-Modifying; SA: Static Attack; DA: Dynamic Attack; OA: Online Attack.}
	\vspace{-0.1in}
	\label{tab:relatedWorks}
	\begin{adjustbox}{max width=0.8\linewidth}
		\setlength{\tabcolsep}{0.36em}
		\begin{tabular}{rcc|cccc}%cccccc}%cccccccccccc}
			\toprule 
			
%			&\rota{Object-Modifying}
%			&\rota{Sensor-Modifying}
%			&\rota{Static Attack}
%			&\rota{Dynamic Attack}
%			&\rota{Real-Time Attack}
%			\tabularnewline
			
			&OM
			&SM
			&SA
			&DA
			&OA
			\tabularnewline
			
			% Eykholt \etal\cite{eykholt2018robust} (2018) & $\Tdot$ & $\Twdot$ & $\Tdot$ & $\Twdot$ & $\Twdot$
			% \tabularnewline
			
			% Athalye \etal\ \cite{athalye2018synthesizing} (2018) & $\Tdot$ & $\Twdot$ & $\Tdot$ & $\Twdot$ &  $\Twdot$ 
			% \tabularnewline
			
			Li \etal\cite{li2019adversarial} (2019) & $\Twdot$ & $\Tdot$ & $\Tdot$ & $\Twdot$ & $\Twdot$ 
			\tabularnewline
			
			Nassi \etal\cite{nassi2020phantom} (2020) & $\Tdot$ & $\Twdot$ & $\Twdot$ & $\Tdot$ & $\LEFTcircle$
			\tabularnewline
			
			Lovisotto \etal\cite{lovisotto2020slap} (2020) & $\Tdot$ & $\Twdot$ & $\Tdot$ & $\LEFTcircle$ & $\Twdot$ 
			\tabularnewline

   			Zolfi \etal\cite{zolfi2021translucent} (2021) & $\Twdot$ & $\Tdot$ & $\Tdot$ & $\Twdot$ & $\Twdot$ 
			\tabularnewline

               Wang \etal\cite{wang2021daedalus} (2021) & $\Tdot$ & $\Twdot$ & $\Tdot$ & $\Twdot$ & $\Twdot$ 
               \tabularnewline

               Wang \etal\cite{wang2021dual} (2021) & $\Tdot$ & $\Twdot$ & $\Tdot$ & $\Twdot$ & $\Twdot$ 
               \tabularnewline

               % Ji \etal\cite{ji2021poltergeist} (2021) & $\Twdot$ & $\Tdot$ & $\Tdot$ & $\Tdot$ & $\LEFTcircle$ 
               % \tabularnewline   

               % Kohler \etal\cite{kohler2021they} (2021) & $\Twdot$ & $\Tdot$ & $\Tdot$ & $\Twdot$ & $\LEFTcircle$ 
               % \tabularnewline  
			
			Jia \etal\cite{jia2022fooling} (2022) & $\Tdot$ & $\Twdot$ & $\Tdot$ & $\LEFTcircle$ & $\Twdot$ 
			\tabularnewline

			\name (this work) & $\Twdot$ & $\Tdot$ & $\Tdot$ & $\Tdot$ & $\Tdot$ 	      
			
			\tabularnewline
			\bottomrule
		\end{tabular}
	\end{adjustbox}
	\vspace{-0.15in}
\end{table}
%\matt{Reformulation of table 1; May need to modify column titles to send the right message. Also, might remove Nassi (very different system \& threat model)}

\section{Preliminaries}
\label{sec:preliminaries}
In this section, we will describe common object perception pipelines in safety-critical, camera-based autonomous systems. 
% and describe the motivating use cases for the rest of the paper. 
We then provide an overview of the different classes of physical AEs presented in Table~\ref{tab:relatedWorks}. Finally, we provide the necessary preliminaries on the transparent display technology used to realize \namenospace. %In particular, \name aims to  We then define our system and attacker model.
% Transition into adversarial example generation --> the physical world is a blackbox

\subsection{Camera-based Object Detection}
%\luis{Add a subsection distinguishing between objectness and classification}
% The sense-to-actuate pipeline for deep learning-enabled, camera-based safety-critical systems such as autonomous vehicles
Deep learning-enabled, camera-based safety-critical systems are typically compartmentalized into perception, planning, and control stages~\cite{pendleton2017perception}. In this paper, we target the perception stage, where a camera captures the light reflecting off scene objects. The camera produces a single frame within a video sequence for a given time interval. Next, the object recognition module behind the optical sensors detects, regionally segments, and classifies the objects within each frame. Finally, the perceived objects are fed into the planning and control stages, which are domain-specific and tied to the underlying applications. Crowd surveillance and analysis~\cite{sreenu2019intelligent} and autonomous navigation~\cite{bagloee2016autonomous} are both examples of such systems. 

\subsection{Physical Adversarial Examples}  
%\luis{This entire subsection is redundant with our description in the intro. We may want to copy and paste paragraphs 2 and 3 of the intro here (including the table), and then shorten/merge the second and third paragraphs of the introduction, removing the table from the intro as well.}
\begin{comment}
\begin{itemize}
    \item One paragraph introducing them (in the context of Table 1)
    \item Transferring digital attacks to physical space
    \item Targeted vs. untargeted attacks \luis{This could be subpoint of digital attacks}
    \item sensor-based vs object-level attacks
    \item Static vs Dynamic attacks
    \item Detection
\end{itemize}
\end{comment}
A substantial body of research focuses on adversarial examples (AEs) in the digital domain~\cite{wang2019security}. 
% where the adversarial perturbations are applied digitally as mathematical operations between two matrices. 
Recent research has emerged where AEs are crafted in the physical domain to demonstrate the practicality of these attacks in the real world. In order to account for dynamic physical environmental conditions, many existing physical attacks employ Expectation over Transformation (EOT) \cite{athalye2018synthesizing} when generating adversarial perturbations. EOT is a data augmentation technique where data is synthesized using various transformations naturally found in a physical environment, such as varying lighting and camera angle. The adversarial perturbation pipeline highly depends on the placement of the perturbation relative to the victim sensor as well as the target perceived object.

As shown in Table~\ref{tab:relatedWorks}, we categorize physical domain attacks based on the location of the perturbation, i.e., object-level modification (OM) versus sensor-level modification (SM), as well as the dynamicity of the attack, i.e., static attacks (SA) calculated ahead of time versus dynamic attacks (DA) that can change at runtime. Additionally, we distinguish online attacks (OA) that generate dynamic perturbations based on the current context of the attack. %on the sensor)for physical domain attacks can be classified along two axes: object-level to sensor-level attacks, and static to dynamic attacks. 
Early static approaches targeted specific, individual objects~\cite{eykholt2018robust,chen2018shapeshifter,athalye2018synthesizing,thys2019fooling,zhao2019seeing,wu2020making, jia2022fooling, wang2021dual, wang2021daedalus}. These attacks change the appearance of a target object by attaching carefully crafted adversarial ``patches" (e.g., stickers or posters) to the object. However, these static attacks have the disadvantage that the patches cannot be changed once applied.
%A vast body of research has been studying adversarial examples in the digital domain~\cite{wang2019security}, i.e., the adversarial perturbations are applied digitally as mathematical operations between two matrices. However, recent research has emerged where adversarial examples are crafted in the physical domain. As shown in Table~\ref{tab:attack-cats}, existing approaches can be classified along two axes: object-level to sensor-level attacks and static to dynamic attacks. The very early works are static attacks targeting specific, individual objects~\cite{}. These attacks apply carefully crafted adversarial patches \eric{is ``patches'' the best word here?}\yi{The term ``patch'' is used in prior work} (e.g., stickers or posters) to the object. These static attacks have the disadvantage that once applied, the patches cannot be changed. For example, if a stop sign is attacked by one of these adversarial patches, it will be perceived by all the vehicles passing by. \eric{replace ``vehicles'' with ``vehicles and agents''?} This is equivalent to removing the sign or replacing the sign forever. 
% \eric{remove ``basically''. too imprecise} 

More recent efforts focus on enabling dynamic adversarial attacks to increase stealthiness or situational awareness. For example, an adversarial perturbation would only be displayed when a target vehicle is approaching a target street sign. Similarly, different perturbations could be generated for targeting different street signs. Lovisotto et al. \cite{lovisotto2020slap} propose using a projector to cast adversarial perturbations onto the object. Such an attack works well for dynamically generating perturbations for specific objects, like a stop sign at a particular intersection. Recent attacks also proposed sensor-first approaches. Instead of altering objects in a scene, they focus on what the optical sensor perceives. In the work of Li et al.~\cite{li2019adversarial} and Zolfi et al.~\cite{zolfi2021translucent}, an adversary prints adversarial perturbations on a transparent paper and attaches the paper to a victim's camera. In this way, the adversarial perturbation follows the victim camera continuously. However, the sticker perturbations are static and do not have the advantages of stealthiness and situational awareness associated with dynamic attacks. In this paper, we aim to bridge the gap between both approaches by proposing a dynamic, sensor-first attack framework. We now describe the transparent mediums that enable such a framework.

\subsection{Transparent Displays}
%\luis{This subsection should be about mediums we can use. Probably can use static vs dynamic}

%\luis{TODO (MAYBE): add a small figure here showing the components of an AR display and how we intend to use it.}
A dynamic, sensor-first attack hinges on the availability of an appropriate attack medium. For camera-based sensing systems, our desired attack vector is light. We seek a medium that will project directly into a camera's optics. The ideal medium is a transparent display that 1) supports dynamic perturbations--as in the projector-based attacks~\cite{hoory2020dynamic,lovisotto2020slap,nichols2018projecting,nassi2020phantom}, 2) can generate perturbations based on the context of a target object ~\cite{nassi2020phantom}, and 3) resides on or near the camera lens such that all light perceived by the camera is transmitted through the medium--as in the case of adversarial camera stickers~\cite{li2019adversarial}. A variety of transparent displays are available as consumer products, with a majority catered to augmented reality (AR) applications such as AR glasses or heads-up displays (HUDs). These products display visual objects and auxiliary information within the user's field of view for various applications, including gaming~\cite{thomas2012survey}, driving assistance~\cite{abdi2015vehicle}, surgical assistance~\cite{vavra2017recent}, and construction safety~\cite{li2018critical}. 

At their core, existing AR solutions consist of two main components: a light source and a holographic combiner (i.e., a specialized lens) that renders images from the projected light into the user's field of view. Precise positioning of the combiner and projection source results in the user simultaneously viewing a scene along with the overlaid projection. Thus, AR technology provides an ideal transparent medium to project dynamic perturbations into the field-of-view of a target camera--serving as a premise for \namenospace's design.
\section{Threat Model}\label{sec:overview}
This section describes the system and attacker models we consider for the \name framework, including a description of the use-cases we focus on and overall attack goals. %We then provide an overview of the design challenges and goals.

\subsection{System Model}
%This work focuses on safety-critical autonomous systems that rely on camera-based perception. We generally consider perception-to-actuation scenarios composed of several steps. Initially, the scene from the optical sensor is converted into a digital image. Next, a deep learning-based perception model is used to perform both object detection and classification on the image. Specifically, perceived objects (e.g., a pedestrian or stop sign) are first detected and isolated within the image, then fed into a trained classifier (e.g., a traffic sign classifier). Finally, the classifier output informs the autonomous system's actions. Example use cases can be crowd surveillance where object recognition is used on camera feeds to detect human subjects of interest, or autonomous navigation, where a vehicle uses cameras to navigate through its environment.%, and security camera scenarios, where facial recognition is used for authentication. 
%\eric{this is repeated from Section 2.a}
This work focuses on safety-critical autonomous systems that rely on camera-based perception. Examples include crowd surveillance where object recognition is used on camera feeds to detect human subjects of interest, or autonomous navigation, where a vehicle uses cameras to navigate through its environment. In general, these perception-to-acutation pipelines are composed of several steps. Initially, the scene from the optical sensor is converted into a digital image. Next, a deep learning-based perception model is used to perform both object detection and classification on the image. 
% We test several well-known model architectures, like the Resnet-50 classifier and the Faster-RCNN and YOLO-V3 object detectors, in the common context of traffic sign recognition. The resulting model outputs inform the autonomous system's actions.

% Moved from attacker goals
We assume several protection mechanisms on the autonomous system to preclude basic attacks. Autonomous vehicles are often equipped with obstruction detection mechanisms and will disable capabilities such as autonomous steering when an obstruction is detected~\cite{souman2021human}. Additionally, some applications have human-in-the-loop control~\cite{sreenu2019intelligent} or may review footage in post-incident analyses, detecting attacks trying to directly fool the classifier (e.g., with overlaid images).

\subsection{Attacker Model}
\begin{comment}
    \item An attacker prepare universal adversarial perturbations (UAP) in advance.
    \item Alternatively, an attacker places a discreet secondary sensor observing the same space as the target sensor
    \item The attacker places a transparent displaying gadget in front of the target sensor, The device can display an arbitrary pattern controlled by the attack.
    %\autoref{fig:setup-diagram} shows the pico projector setup.
    \item Although the attacker does not have access to any software component of the target system, we assume the attacker has knowledge of the model. This model is consistent with prior works, and transferrability/generalizability of attacks is outside of the scope of the paper.
    \item Basically we assume a white box attack scenario, i.e., the attacker has access to the internal states (e.g. weights, gradients w.r.t. an input image, etc) of the model. 
    \item 
\end{comment}

% As depicted in Figure~\ref{fig:attack-overview}, 
We assume an adversary attaches a portable and low-profile gadget, consisting of an adversary-controlled transparent display and auxiliary sensor, in front of the victim's optical sensor or camera. We assume that an auxiliary sensor can sense when a target object enters the victim camera's field of view. The auxiliary sensor can take on many forms depending on the target application's real-time requirements and the attacker's resource constraints. For instance, the auxiliary sensor can be another camera attached alongside the victim camera with on-device object detection, or the sensor could be a GPS tracking device coupled with a map of traffic signs throughout the city. Alternatively, the ``sensor" could be the attacker remotely observing the victim camera's movement while controlling the transparent gadget. The auxiliary sensor informs the transparent display gadget which adversarial perturbation to display. We assume that the adversary can prepare adversarial perturbations in advance for each target class or that the real-time requirements for the target application allow the controller to generate perturbations at runtime\footnote{We note that current commercial off-the-shelf embedded technology may not be able to run real-time perturbation generation, so we assume that the adversary can prepare perturbations in advance as a practical workaround.}. %\matt{This assumption seems shaky when taken as a whole; I think that it might be better to word this something like: we can concede that current COTS embedded tech may not be strong enough to run real-time perturbation generation, so (as a practical workaround) we assume that an adversary can prepare perturbations in advance.}
%Finally, we assume that the transparent display does not degrade the performance of the target application when no light is projected onto the display, i.e., when there are no perturbations. \matt{Do we need to assume this? We have ablation studies showing no loss in accuracy when nothing is projected}  
%Unlike prior dynamic physical attack frameworks, which make changes to specific target objects, we do not assume to know the predetermined path of the target application so that the attacker can situationally adapt perturbations given the auxiliary sensor data. 
Previous physical attack frameworks require prior knowledge of a victim's intended route in order to physically modify objects (e.g. with stickers) before they are in view, limiting dynamicity. Using information from an auxiliary sensor (e.g. a GPS),  we can situationally adapt perturbations to attack the current object in view.

\noindent\textbf{Attacker goals.} The attacker's primary objective is to fool the victim application's recognition module, leading to erroneous object predictions which compromise safety, all while evading detection. This means that 1) employing denial-of-service attacks (e.g. applying opaque obstructions) or directly presenting adversarial images (e.g. an incorrect traffic sign) will be ineffective; 2) perturbing system optics during other states of normal operation is also undesirable, as it increases risk of detection; and 3) the attack should be able function under varied environmental conditions, which especially important for light-based attacks. \name is designed to achieve this goal while navigating the aforementioned constraints, generating dynamic perturbations optimized to look like chromatic aberrations or distortions of the camera sensor.  

\begin{figure*}[]
\centering
\includegraphics[width=0.8\linewidth]{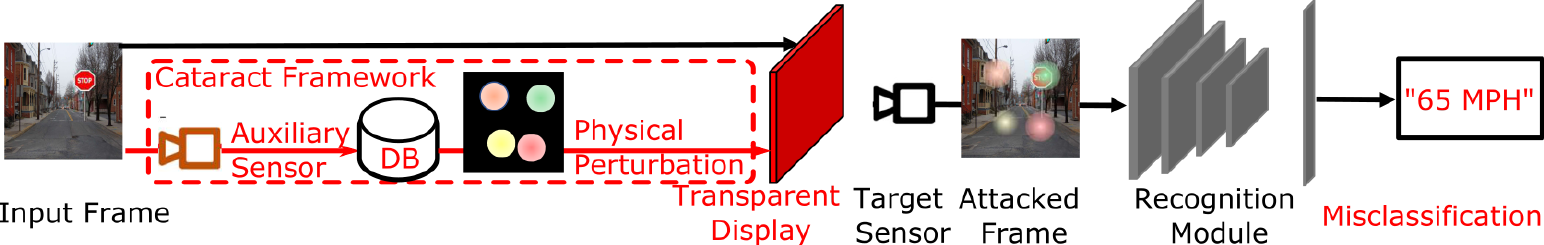}
\caption{Overview of \name framework. An attacker uses a transparent display and an auxiliary sensor to implement an optical man-in-the-middle attack against a target sensor. The attacker first uses the auxiliary sensor to determine the target object for an input frame. \name then looks up a pretrained digital perturbation to create an unsafe misclassification. %\namenospace's Tnet transforms the digital perturbation into a physical perturbation, i.e., the projection of the digital perturbation onto the transparent display. 
In this example, a ``Stop" sign is being perceived as a speed-limit sign by the victim system. 
%\yi{it is digital perturbation }
}
\label{fig:diagram}
\end{figure*}
\section{Methodology}\label{sec:design} 
In this section, we first formalize the research challenges for dynamically generating adversarial perturbations at the sensor-level. We then briefly provide an overview of how our design aims to address each research challenge. Finally, we describe the methodology of each design component for our proposed attack framework. 

\begin{figure}
\centering
\begin{subfigure}{.4\linewidth}
  \centering
  \includegraphics[width=.6\linewidth]{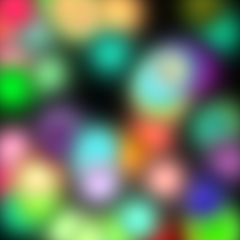}
  \caption{Digital perturbation.}
  \label{fig:p_digital}
\end{subfigure}%
\begin{subfigure}{.4\linewidth}
  \centering
  \includegraphics[width=.6\linewidth]{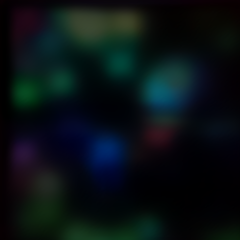}
  \caption{Physical perturbation.}
  \label{fig:p_physical}
\end{subfigure}
\caption{A pair of digital and physical perturbations. The physical perturbation looks different from the digital perturbation due to the characteristics of the display and sensing pipeline.
\yi{present these images with an actually stop sign background}}
\label{fig:perturbation_pair}
\vspace{-0.2in}
\end{figure}

\subsection{Problem Formalization}
%\matt{todo: check for double parenthesized abbreviations and abbreviations that are only used once (e.g. CDTF defined twice with two different references)}
\noindent\textbf{Research Challenge \#1: Designing perturbations robust to optical transformations.} Directly applying an adversarial perturbation computed in the digital domain fails because it does not account for the mutations introduced by the camera-display transfer function (CDTF)~\cite{wengrowski2016optimal}, i.e., the transformation of a pixel projected in the physical world to the pixel captured by the camera. In our case, an adversarial perturbation is displayed and perceived by the camera, going through a set of transformations as depicted in Figure~\ref{fig:perturbation_pair}. The color spectrum emitted by an electronic display may not match the color spectral sensitivity function of the camera~\cite{wengrowski2016optimal}, leading to less color-robust perturbations. 
% Prior work~\cite{lovisotto2020slap, wengrowski2019light} addressed this challenge by using a neural network to learn the color transformation. 
Moreover, because the transparent display is in close proximity to the victim camera, the displayed perturbation is blurred in the perceived image due to the Bokeh effect~\cite{ignatov2020rendering}. The image formation pipeline processes within the camera converts the light received by the image sensor to a final digital image that is then affected by exposure, the sensor's sensitivity (ISO), and contrast transforms. %It would be too complicated to model each of these transformations one by one, so we employ an end-to-end approach in our attack~\cite{wengrowski2019light}.

\noindent\textbf{Research Challenge \#2: Designing perturbations robust to environmental conditions.} Multiple factors must be considered to make an adversarial perturbation robust to dynamically changing conditions in the physical environment. For instance, the camera's perspective, distance, and rotation relative to a target object will all impact the efficacy of a perturbation. The scene's environmental background and ambient illuminance also play a vital role in object recognition. %We propose an automated data collection framework that incorporates these factors during the perturbation training and generation process. This approach is analogous to data augmentation techniques employed by prior dynamic approaches~\cite{lovisotto2020slap}. However, in our case, we explicitly account for the impact of environmental factors to enhance situational awareness in downstream perturbation tasks. For instance, the training process may find that certain classes of adversarial examples only work for specific illuminance ranges. 

\noindent\textbf{Research Challenge \#3: Efficiently generating perturbations for real-time applications.} In addition to optical and physical mutations, the transient nature of real-time applications introduces a significant challenge to AE generation.
Environmental conditions can change significantly within a short period of time.
%The variance in an environment may be significant across multiple frames. 
Because generating a perturbation incurs latency in the attack pipeline, attack perturbations should be generated to perform well across this environmental variance, rather than attempt to optimize for a single frame. %Thus, Cataract adopts universal adversarial perturbation (UAP) models~\cite{moosavi2017universal} to generate perturbations that will be robust against variance across multiple time frames for a given scene.

\begin{figure*}[]
\centering
\includegraphics[width=0.8\linewidth]{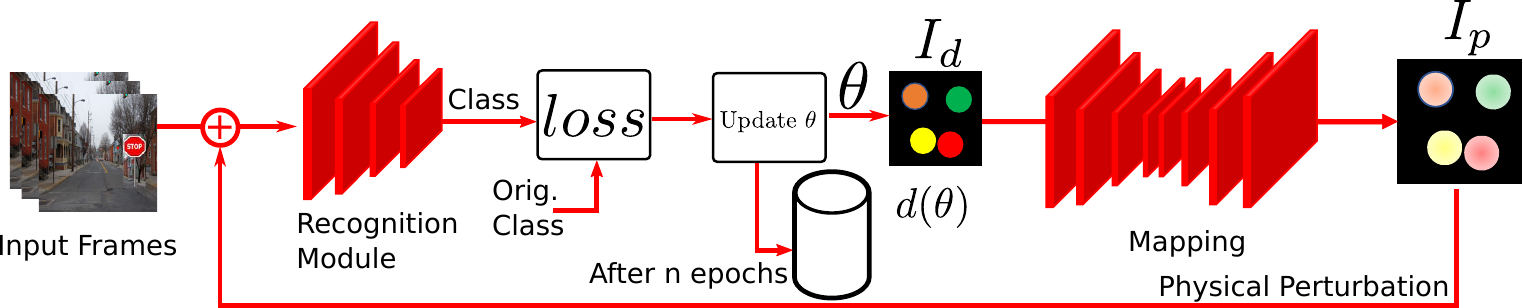}
\caption{Overview of training process for physical perturbations in the \name framework. \name   generates a digital perturbation to create an unsafe misclassification. \namenospace's digital-to-physical mapping then transforms the digital perturbation into a physical perturbation, i.e., the projection of the digital perturbation onto the transparent display. %\nils{ please update caption}
%\eric{Perhaps we should rearrange this figure to look more like an man-in-the-middle attack? Right now it just looks like a series of neural nets. Some color would be helpful here too.}
%\eric{Need a better font. Can we replace the "65 MPH" text with the image of a 65 MPH sign? Would also be good to have an image of a Stop Sign (the correct class). Otherwise, this figure is not clear at first glance.}
%\luis{We may want to remove the auxiliary sensor and recognition model from the Cataract framework box. One solution for the "Recognition Module" in the cataract framework could be to have a more general term to refer to the mechanism that determines what is the correct digital perturbation to generate. We may also need to add the transparent display within the cataract framework box.}
%\luis{We may want to explicitly refer to TNet}
}
\label{fig:training}
\vspace{-0.1in}
\end{figure*}

\noindent\textbf{Design Overview.} Figure~\ref{fig:training} depicts an overview of the \name framework. To address the digital-to-physical mapping of \textbf{Research Challenge \#1} and \textbf{Research Challenge \#2}, we employ an end-to-end deep learning approach for modeling the optical transformations (we provide details in Section~\ref{sec:modeling}). We incorporate environmental background noise and illuminance variations into an automated data collection framework for perturbation training and generation (details in Section~\ref{sec:crafting-adversarial}). Finally, to address the real-time dynamics discussed in \textbf{Research Challenge \#3}, we adopt a universal adversarial perturbation (UAP) model~\cite{moosavi2017universal} that is robust against variance across multiple time frames for a given scene. Moreover, we leverage semantic information to efficiently generate this UAP (detailed in Section~\ref{sec:crafting-adversarial}).

\subsection{Modeling Digital-to-Physical Perturbations}\label{sec:modeling}

% The normal operation of a recognition model can be described as 
% \begin{equation}
%     y = F(x)
% \end{equation}
% where $x$ is a frame of the video stream perceived by the victim camera. $y$ is the true label of the perceived object. $F(\cdot)$ is the recognition model. 
Under the additive adversarial perturbation model, the effect of a physical adversarial perturbation can be abstracted as 
\begin{equation}
    y' = F(x+I_p),\quad y\neq y'
\end{equation}
where $x$ is a frame of the video stream perceived by the victim camera. $I_p$ is a physical adversarial perturbation. $F(\cdot)$ is the recognition model. $y$ is the true label of the perceived object. $y'$ is an intentionally misclassified label. $I_p$ can be produced in different ways such as printed on a sticker and attached to the object \cite{eykholt2018robust}, printed on a transparent paper and attached to the camera sensor \cite{li2019adversarial} or projected onto the object using a projector \cite{lovisotto2020slap}. In our attack, $I_p$ is presented on a transparent display and attached to the camera sensor. The camera sensor can see through the display.

Applying adversarial perturbation computed in the digital domain fails due to the mutations/transformations introduced by the CDTF~\cite{wengrowski2016optimal}. %camera-display transfer function (CDTF) \cite{wengrowski2017reading}. 
Such mutations are caused by the characteristics of the displaying and camera sensing pipeline. On the transparent display side, the optics of the holographic combiner and projection source such as color spectrum compression and shifting, distortion will already produce an image different from the source image. On the sensor side, the optics of the camera, such as focal length, lens distortion, exposure, and white balance also affect the final perceived image. Therefore, in order to produce effective adversarial perturbations in the physical domain, it is necessary to understand and consider these mutations in the design loop. There is prior effort attempting to model these mutations. Lovisotto et al.~\cite{lovisotto2020slap} used a multi-layer perceptron (MLP) to model the color spectrum mapping between intended and perceived colors given the color of the projection surface. while this mapping model works fine for their attack purpose, it requires discretizing the color space to reduce complexity, which could lead to imprecision. Besides, to achieve a more precise digital-to-physical mapping and thus a more powerful attack, other mutating factors mentioned above also need to be considered. In this paper, we propose establishing a digital-to-physical mapping that takes all these mutations into consideration. 

% To model these mutations, we propose establishing a mapping between an adversarial perturbation computed in the digital domain and its corresponding physical perturbation.

\noindent\textbf{Digital-to-physical mapping.} Given a pair of digital and physical perturbation image $I_d$ and $I_p$, we wish to find a mapping $T(\cdot)$ such that 
\begin{equation}
    I_p = T_{\Theta, D, C, E}(I_d)
\end{equation}
where $\Theta$ represents the domain-constraint parameters of the perturbation generation (e.g., the size, shape, and color constraints of a perturbation), $D$ represents the transparent display, $C$ denotes the camera, and $E$ denotes  environmental parameters (e.g., ambient light). It would be too complicated to model each component, i.e., aforementioned mutations, individually. Therefore, we take a data-driven approach. We employ a convolutional neural network to approximate $T(\cdot)$.

% take care of the display $D$ and the camera $C$ (denoted as $T_{D, C}(\cdot)$) and we take care of the environment $E$ on top of it (denoted as $T_E(\cdot)$)
% \begin{equation}
%     I_p = T_E(T_{D, C}(I_d)).
%     \label{eq:mapping}
% \end{equation}
% % Since $T(\cdot)$'s job is to approximates the mutations/transformations introduced by the CDTF, we call it TNet. 
% 

\noindent\textbf{Fitting the model.} $T(\cdot)$ is fitted with pairs of digital and physical perturbations. We generate a dataset of $I_d$ and $I_p$ pairs as the training set. For each pair, we first randomly generate and display $I_d$ on the transparent display. We then take two images, one with $I_d$ turned on (mutated image) and one with $I_d$ turned off (the background). We derive $I_p$ by subtracting the background from the mutated image. When we collect data for fitting $T(\cdot)$, no object is present in the scene. 

As for generating $I_d$, we use highly structured patterns consisting of dots with different colors. Extending the idea of using small dots to create adversarial perturbations in \cite{li2019adversarial}, we define a dot-based adversarial perturbation generation function $I_{d, \theta}$. The parameters contained in $\theta$ describe the dots, and are defined as 1) $c(i^{(k)}, j^{(k)})$ - center coordinates of the $k$th dot; 2) $r_k$ - radius of the $k$th dot; 3) $\gamma_k$ - RGB color of the $k$th dot; 4) $\alpha_{\textnormal{max}}$ - maximum alpha blending value; 5) $\beta$ - exponential dropoff of alpha value; 6) $n$ - maximum number of dots.
% \begin{itemize}
%     \item $c(i^{(k)}, j^{(k)})$ - center coordinates of the $k$th dot
%     \item $r_k$ - radius of the $k$th dot
%     \item $\gamma_k$ - RGB color of the $k$th dot
%     \item $\alpha_{\textnormal{max}}$ - maximum alpha blending value
%     \item $\beta$ - exponential dropoff of alpha value
%     \item $n$ - maximum number of dots.
% \end{itemize}
$I_d$ can be computed as
\begin{equation}
    I_d(i,j) = \sum_{k=1}^{n}\alpha_{\textnormal{max}}e^{-\beta \cdot d_k}\cdot\gamma_k
\end{equation}
where $d_k(i,j)$ is the distance of the pixel $(i,j)$ with respect to the center of the $k$th dot
\begin{equation}
    d_k(i,j) = \frac{(i-i^{(k)})^2+(j-j^{(k)})^2}{r_k^2}. \\
\end{equation}
$I_d$ superimposes $n$ dots together to create a pattern with various colors and shapes. \autoref{fig:p_digital} shows the example of a digital perturbation. The number of dots in an adversarial perturbation affects its capacity; an adversarial perturbation with fewer dots is less powerful than one with more dots. A perturbation with more dots can create more complex patterns and, thus, can alter the frames and the prediction of the target recognition model more easily. The advantages of using such highly structured patterns are 1) small modulations to pixel values, as in pixel level adversarial perturbations, do not survive when displayed and perceived on an object; 2) It would require significantly more data in the training set in order to ensure generalizability, i.e., to enable $T(\cdot)$ to output physical perturbations based on unseen digital perturbations.

To fit $T(\cdot)$, we want the output of $T(\cdot)$ to be as similar as possible to the ground truth physical perturbation $I_p$. To achieve this goal, we combine both mean square error (MSE) and Learned Perceptual Image Patch Similarity (LPIPS)--a metric optimized for perceptual loss \cite{zhang2018perceptual}, to form the loss function $l$
\begin{equation}
    l = (1 - a) * \textnormal{MSE}(T(I_d), I_p) + a * \textnormal{LPIPS}(T(I_d), I_p)
    \label{eq:loss}
\end{equation}
where $a$ controls the ratio of the two terms. The MSE term encourages the colors to be as close as possible, the LPIPS loss term emphasizes maintaining the structural patterns.

\subsection{Crafting Adversarial Perturbations}\label{sec:crafting-adversarial}

In this section, we describe how we design a digital and physical co-optimization framework %, which refer to as \textit{TNet},\yi{do we want to bring up the term TNet at this point?} 
to find the adversarial perturbation that can evade the target recognition model.

\noindent\textbf{Attack Objective.} 
% In adversarial example (AE) generation research, a unique AE can be computed for each individual image under attack, or one AE can be computed to work for a set of images. This is called a universal adversarial perturbation (UAP). 
As is mentioned in research challenge \#3, UAP fits better in our attack scenario: an adversary prepares adversarial perturbations for the classes of images they intend to attack beforehand.
% \eric{why does UAP fits better in our attack scenario?} 
%For instance, recognize stop signs as some other traffic signs or recognize the face of a burglar as the house owner.
Each one of these UAPs can cause an incorrect prediction for all the instances of the object in the corresponding class. The adversary then attaches the attack gadget to the victim camera, and remotely controls which UAP to display on the gadget. Thus, we define our attack objective as
\begin{equation}
    \arg\min_{\theta}\quad \mathbf{E}_{x\sim D(x|y)}l[(F(x+I_p))]
\label{eq:objective}
\end{equation}
where $\theta$ represents the free parameters of the perturbation generation function (e.g., $c$ and $\gamma$).  $l$ is the loss function of the attack. A sample $x$ is drawn from a distribution $D(x|y)$ (e.g., the class of all stop signs). Depending on the type of attacks the adversary wants to perform and the target recognition model, $l$ can take various forms. For instance, for an untargeted attack on a traffic sign classifier (i.e., to classify a frame into any class other than the true class), $l$ is the reciprocal of the cross entropy loss with respect to the true class. For an untargeted attack on a traffic sign detector, $l$ is the average class score of all the predicted bounding boxes with respect to the class under attack (see Section~\ref{sec:prototype} for details). 
%\luis{TODO: Yi, can you address Saman's comments and clarify the above parameters?}

\noindent\textbf{Optimizing a Successful Attack.} As each module in our attack pipeline is differentiable, it is theoretically possible to approach a solution of \autoref{eq:objective} using gradient descent. However in practice, we found the effectiveness of an optimized perturbation to be highly sensitive to the initialization of free parameters, namely, $c$ and $\gamma$. Also, as shown in~\cite{li2019adversarial}, the gradients with respect to the free parameters present a highly non-convex loss surface. Therefore, we first find a good initialization using a coarse grained greedy block coordinate descent search. We then apply fine-grained gradient descent using this initialization. Specifically, we split a perturbation into blocks of the same size. The center of these blocks are the candidate locations of the dots $c$. We also discretize the RGB color space to obtain a fixed set of candidate colors. We then optimize for one dot at a time. For each dot, we try all the candidate locations and colors and pick the one that gives the maximum loss (to have a higher chance of misleading the recognition model). We repeat this process until convergence. Next, using the dots computed above as initialization, we iteratively compute the gradients of the loss with respect to the free parameters and extract the sign of the gradient. We then add a small step in the direction of the sign. Detailed steps of our optimization are shown in Algorithm~\ref{algo:attack}. Note during the optimization of the attack the weights of $T(\cdot)$ remain fixed.
% \eric{good}

\noindent\textbf{Serving the Attack.} Attack optimization is performed in advance for the objects under attack. The generated perturbations are stored in a database. At runtime, the attacker first uses the auxiliary sensor to determine the context of the runtime environment (e.g., a target object entering the scene). \name then displays a pre-generated perturbation fetched from the database to create an unsafe classification.

\begin{algorithm}%[H]
\DontPrintSemicolon
  \KwInput{Images under attack $X=\{x1,x2,...x_n\}$, attack objective $l$, maximum number of iterations \textit{maxiter}, perturbation generation function $\theta$}
  \KwOutput{Digital perturbation $I_d$}

%   \KwData{Testing set $x$}
%   $\sum_{i=1}^{\infty} := 0$ \tcp*{this is a comment}
%   \tcc{Now this is an if...else conditional loop}
%   \If{Condition 1}
%     {
%         Do something    \tcp*{this is another comment}
%         \If{sub-Condition}
%         {Do a lot}
%     }
%     \ElseIf{Condition 2}
%     {
%     	Do Otherwise \;
%         \tcc{Now this is a for loop}
%         \For{sequence}    
%         { 
%         	loop instructions
%         }
%     }
%     \Else
%     {
%     	Do the rest
%     }
    % \tcc{Now this is a While loop}

    Initialize $\theta$ with coarse grained greedy coordinate descent search \\
    \While{\textsc{\textbf{not}} converge \& iter $\leq$ maxiter}
   {
        1. Compute digital perturbation $I_d$ \\
        2. Compute physical perturbation $I_p=T(I_d)$ \\
        % 3. Estimate and adjust the brightness of $T_{D,C}(I_d)$ based on the ambient light to get $I_p$ \\
        3. Apply $I_p$ to X  \\
        4. Query the victim model and compute the attack objective $l$ \\
        5. Perform a step of gradient descent to update $\theta$
   }
\caption{Physical AE Generation}
\label{algo:attack}

\end{algorithm}

\noindent\textbf{Robust Physical Adversarial Examples} 
To make an adversarial perturbation robust against dynamic physical environmental conditions, we consider the following factors when preparing images for the adversarial perturbation generation process:
\begin{itemize}
\item \textbf{Background.} The context of the background plays a role in recognizing an object. We use images with various backgrounds within the same context.

\item \textbf{Perspective and rotation.} The object might come in a different perspective and rotated in front of the camera. We prepare images with various perspectives and rotations. This is application-specific, e.g., when attacking a traffic sign recognition model, we consider perspectives between $-30^{\circ}$ and $30^{\circ}$, with rotations between $-5^{\circ}$ and $5^{\circ}$.

\item \textbf{Distance.} The object might also be located at various distances from the camera. To account for this, we consider various sizes of the object.
% \item \textbf{Illuminance.} To consider different environmental light conditions, we apply brightness transformation to the simulated attacked images.

\item \textbf{Illuminance.} To account for different light conditions, we vary the ambient light of the environment when collecting images. Specifically, we consider an illuminance ranging from 30 lux to 3000 lux.
\end{itemize}

% To make an adversarial perturbation work under various environmental noise, e.g., viewing angles, distance, and light conditions, etc., we propose to use Expectation over Transformation (EOT), a common method in physical AE generation. EOT works by synthesizing training images using a set of transformations that represents various environmental noise. These images augment the training dataset used for computing an adversarial perturbation. In this way, the generated adversarial perturbations can be more robust. 

% We mainly consider the following transformations
% \begin{itemize}
%     \item To deal with this, we consider the following brightness transformation
%     \begin{equation}
%         g(i, j) = a f(i, j) + b
%     \end{equation}
%     where $g(i, j)$ and $f(i, j)$ are source and target image pixels respectively. We 
%     \item The object under attack might be presented to the victim camera in various angles. To account for this, we first crop the object from the image. We then apply perspective transformation to the object. Finally, we attach the object back to a background image. 
%     \item The object under attack might also be presented to the victim camera in different distances. Following similar approach as angles, we change the size of the object.
% \end{itemize}

\noindent\textbf{Dynamic Adversarial Attack.} One major advantage of our proposed attack compared to prior static attacks is that in our attack approach, the adversary can change the perturbation based upon different scenarios. For example, in a traffic sign recognition use case, if a car is following a particular route, we can assume there are several different safety-critical traffic signs along the route. 
% \eric{a bunch? how about ``several''?} 
To maximize the damage of the attack, The adversary wants to successfully attack as many critical traffic signs as possible. In the static attack setting, the perturbation cannot be changed, and the adversary can only compute a single UAP to accommodate all different kinds of traffic signs. In our dynamic attack setting, we can change the perturbation based on external information such as GPS of the car and a traffic sign map~\cite{ertler2020mapillary, vargas2020openstreetmap} to maximize the attack success rate. Moreover, an attacker can enable perturbations only when desired to minimize alerting the vehicle to a camera obstruction or unwanted behavior during normal operation. 
% \eric{``however'' is used too commonly throughout this paper. not every sentence needs a qualifier.}

\yi{should describe how we store computed perturbations in a database and fetch based on auxiliary information}
\section{Implementation and Evaluation}\label{sec:evaluation}
%\matt{todo: replace instances of Cataract with \name}
In this section, we provide details on the implementation and evaluation of the \name attack. First, we describe the design of our neural-based approach to digital-to-physical mapping $T(\cdot)$, which we call \textit{TNet}. We then describe our prototype and experimental setup. Finally, we evaluate each component of \namenospace, including how the attack fares against existing defenses.

\subsection{Digital-to-Physical Mapping via TNet}

%We describe our realization of the digital-to-physical mapping $T(\cdot)$ here. 
% Recall we split the mapping to two steps, $T_{D,C}(\cdot)$ and $T_E(\cdot)$. 
%We employ a neural network model to approximate $T(\cdot)$. %\noindent\textbf{Model Architecture.} 
%It is essential to choose the right architecture for the digital-to-physical mapping $T(\cdot)$. Since both the input and output of $T(\cdot)$ are images, it is intuitive to think of a fully convolutional neural network model. During our early investigation we tried a simple convolutional auto-encoder. However, this architecture cannot learn the digital-to-physical mapping well (outputing physical perturbations looks distinct from ground truth images). Increasing the size of the model didn't help. So we switched to a UNet architecture~\cite{ronneberger2015u}. The multiple skip connections in UNet enables the model to learn from the residuals between feature maps thus converges faster and has better accuracy. \yi{need rephrasing}
To approximate the non-linear mapping between digital-to-physical perturbations, we design and train a feed-forward neural network model called TNet. Beyond boasting good performance accuracy, feed-forward neural networks are also differentiable. So a pre-trained TNet can be included as a component of the training procedure of \namenospace, where differentiation of the mapping function $T(\cdot)$ is essential to the back-propagation step (see Algorithm~\ref{algo:attack}).

The inputs to TNet are RGB images representing digital perturbations, and the outputs are RGB images that illustrate how these digital perturbations will be perceived by the victim camera in the physical world. Two architectural backbones were initially selected for TNet: a fully-convolutional auto-encoder, and the UNet architecture~\cite{ronneberger2015u}. Empirically, we found that the UNet architecture's multiple skip connections enabled the model to converge faster and with better accuracy across a range of tested batch sizes.

\noindent\textbf{Model Training.} To train the mapping, we randomly vary the center and the RGB vector of each dot and generate $10,000$ digital perturbations. We then collect their corresponding physical perturbations using the aforementioned hardware setup. During our initial investigation, we allowed other parameters of the alpha blending model to vary (e.g., $r$, $\alpha_{max}$, and $\gamma$).
With too many free parameters and insufficient training data to cover the combinatorially growing space, TNet cannot learn a robust mapping. Additionally, if the training dataset is not large enough, TNet overfits the training data, and during the attack phase, resulting in a less-precise computed gradient. The optimization will deviate significantly from the correct direction, leading to an ineffective perturbation. Given the data collection bottleneck with a physical setup, we wanted to minimize the reliance on a large dataset. Thus, we fixed all the parameters except the center $c$ and the RGB vector $\gamma$ of each dot. We found that a dataset with $10,000$ pairs was sufficient to train TNet. %To determine the size of our dataset, we started from a very small number of $2,000$ pairs and gradually increased the number and find out $10000$ pairs is sufficient to train a good TNet for our attack.

We used $80\%$ of the image pairs for training and $20\%$ for validation. As mentioned in Section~\ref{sec:design}, the number of dots affects the capacity of our attack: too few dots cannot perturb an image effectively, while effective perturbations with too many dots are difficult to train without a sufficiently large dataset. To show the effect of different number of dots, we consider $10$, $30$ and $50$ dots when training TNet. We set $\alpha$ in \autoref{eq:loss} to be $0.0004$.
% According to \autoref{eq:loss}, we can control the weights of MSE and Perceptual loss by adjusting the $\alpha$. 
We trained TNet for $1,000$ epochs with a batch size of 32 and a learning rate of $0.003$. We used Adam~\cite{kingma2014adam} as the optimizer.

\noindent\textbf{Evaluating TNet.} Besides MSE and LPIPS~\cite{zhang2018perceptual}--which we used to train TNet--we also consider Peak Signal-to-Noise Ratio (PSNR) and the Structural Similarity Index (SSIM) \cite{hore2010image} to measure the accuracy of TNet. PSNR is derived from MSE, and takes into account the peak intensity of the input images to describe the similarity between two images. SSIM is designed based on luminance, contrast, and structure, so a higher SSIM indicates higher similarity in these factors. 

We compared our UNet-based TNet (TNet-UNet) with two baseline models: a CNN Autoencoder-based model (TNet-CNN) and a multi-layer perceptron (MLP)-based model (TNet-MLP), since prior work~\cite{lovisotto2020slap} used an MLP-based approach. 
%We also consider TNet-MLP because an MLP model is also used in prior work~\cite{lovisotto2020slap}.
For both baseline models, we configure the architecture such that the overall number of weights is similar to TNet-UNet.
%TNet-CNN was initially selected but suffered from poor attack accuracy.

% As can be seen from the results, TNet can approximate digital-to-physical transformation very well.
% We also show the comparison between some ground truth physical perturbation (collected) and TNet output in \autoref{fig:tnet_output}.

\begin{figure}
\centering
\includegraphics[width=0.6\linewidth]{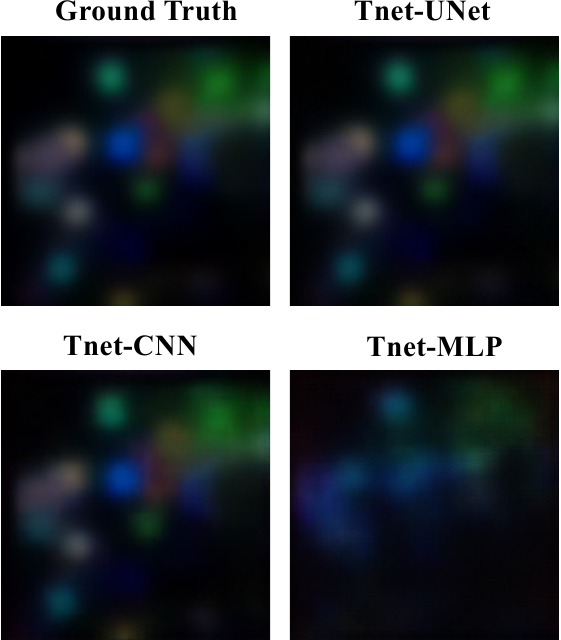}
\caption{Visual comparison between the output of all TNet models and the ground truth of the $50$-dot case. TNet-UNet output approximates the ground truth perturbation best.}
\label{fig:perturbations}
\vspace{-0.15in}
\end{figure}

\autoref{tab:tnet_accuracy} shows various metrics of the trained models on the test dataset. TNet-UNet achieves significantly better results in terms of the validation loss and the four metrics compared to the two baseline models, implying that the UNet architecture is the correct choice to approximate the digital-to-physical mapping with high accuracy. \autoref{fig:perturbations} depicts an example of different TNet models' output and the corresponding ground truth. %We also compare the output of TNet-UNet and its corresponding ground truth in \autoref{fig:perturbations}.
%\yi{visual comparison of all 3 cases.}
%As can be seen from the figures, TNet generates perturbations very close to the actual physical perturbation. 
A more precise physical mapping guarantees more precise gradient values of the pattern generating function in the attack stage, increasing the probability of a successful adversarial perturbation. We select the TNet-UNet architecture for the rest of the paper and refer to it as TNet.

% \begin{table}[]
% \centering
% \begin{tabular}{|c|c|c|c|c|c|}
% \hline
% \multicolumn{1}{|l|}{\# dots} &
%   \multicolumn{1}{l|}{Validation loss} &
%   \multicolumn{1}{l|}{MSE} &
%   \multicolumn{1}{l|}{LPIPS} &
%   \multicolumn{1}{l|}{PSNR} &
%   \multicolumn{1}{l|}{SSIM} \\ \hline
%  10 & 6.457e-3  & 0.965e-3 & 5.393 & 30.8 & 0.887   \\ \hline
%  30 & 9.086e-3 & 1.062e-3 & 8.043 & 29.74 & 0.853     \\ \hline
%  50 & 10.670e-3 & 1.535e-3  & 9.134  & 28.14  & 0.815    \\ \hline
% \end{tabular}
% \caption{Accuracy of TNet. TNet is able to approximate the mapping... 
% % \eric{Which percpetual loss? this doesn't make sense as written. a specicific perceptual loss algorithm should be named and cited. it's probably LPIPS.}
% }
% \label{tab:tnet_accuracy}
% \end{table}

\begin{table}[]
\centering
\caption{Accuracy of TNet models. The total loss of TNet-UNet on the validation set as well as four other image similarity metrics is significantly better than the other two baseline models. There are not fixed ranges for the metrics (of course they all need to be non-negative numbers) except that SSIM is between 0 and 1. However, a smaller MSE/LPIPS or a larger PSNR/SSIM value indicates better similarity between the predicted physical perturbation and the ground truth.}
%\resizebox{\linewidth}{!}{%
\begin{tabular}{cccccc}
\toprule
& Validation loss & MSE    & LPIPS & PSNR  & SSIM \\ \hline
TNet-UNet & \textbf{2.26e-4}         & \textbf{1.95e-4} & \textbf{0.31}  & \textbf{37.17} & \textbf{0.92} \\
TNet-CNN & 2.39e-3 & 1.42e-3 & 9.75 &  28.46 & 0.83  \\
TNet-MLP & 0.01 & 8.74e-3 & 53.76 & 20.58 & 0.58  \\
\bottomrule
\vspace{-0.1in}
\end{tabular}
%}

% \eric{Which percpetual loss? this doesn't make sense as written. a specicific perceptual loss algorithm should be named and cited. it's probably LPIPS.}
% \yi{put results of simple CNN.}}
\label{tab:tnet_accuracy}
\end{table}

\subsection{Prototype and Experimental Setup}
\label{sec:prototype}

\noindent\textbf{Implementation.} The entire adversarial perturbation generation pipeline is implemented in PyTorch~\cite{NEURIPS2019_9015}. We utilize an open-source UNet implementation\footnote{https://github.com/zhixuhao/unet} for the TNet architecture. All training and adversarial perturbation generation jobs are run on an Ubuntu 18.10 computer with $4$ NVIDIA RTX A5000 graphic cards and $132$GB memory. Note in the attack phase the adversary only needs to serve pre-generated perturbations.

% \begin{itemize}
%     \item Transparent Display Setup (include not on AR glasses)
%     \item Dataset we used
%     \item Computation/laptop/camera
%     \item Include figure of setup
%     \item Target models attacked
% \end{itemize}

% \begin{figure}
% \centering
% \includegraphics[width=0.9\linewidth]{\figurepath/setup-diagram}
% \caption{Diagram of our beam splitter-based transparent display gadget. The beam splitter reflects light from the LED display, allowing the object to be perceived at the same time.} 
% % \eric{should we add a 'Beam Splitter' label near the blue box in the graphic?}
% % \yi{added}
% \vspace{-0.2in}
% \label{fig:setup-diagram}
% \end{figure}

%TODO: note, this figure still has Cataract in it
\begin{figure}
\centering
\includegraphics[width=0.9\linewidth]{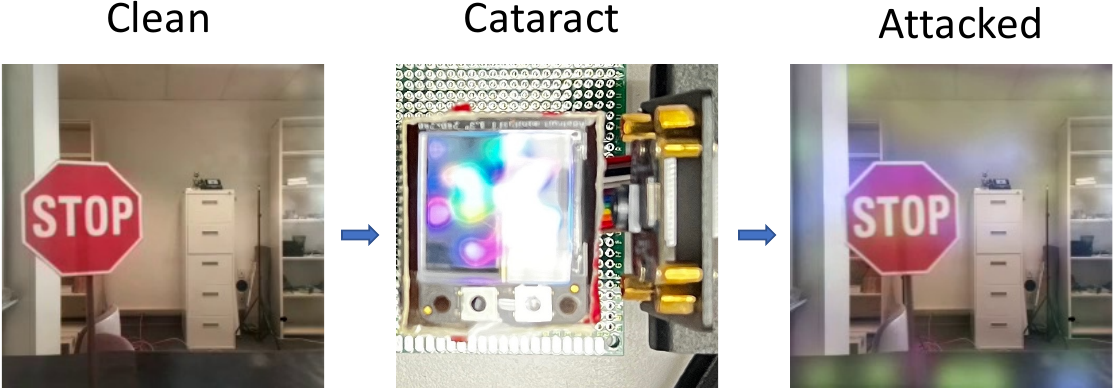}
\caption{Our physical evaluation setup, consisting of a Raspberry Pi victim camera and a low-cost transparent display analogue, consisting of an organic light-emitting diode (OLED) display and beam splitter. Adversarial perturbations are displayed on the screen and reflected by the beam splitter, overlaying them onto the natural scene as observed by the victim camera.
%Our portable and low cost transparent display gadget attached to the victim camera. The gadget consists of a beam splitter, 
% \luis{Add labels; not sure if bottom half of image is necessary}
%\eric{should describe the role of every component in the caption}
%\yi{Take a different figure showing the setup possibly on a Tesla car.}
}
\label{fig:trans-display}
\vspace{-0.1in}
\end{figure}

\noindent\textbf{Transparent Display Gadget.} 
% \autoref{fig:setup-diagram} shows the construction of our transparent display gadget. 
Borrowing the mechanism behind AR glasses and HUDs, we attach a $50:50$ ratio beam splitter to an Adafruit Mini PiTFT $1.3$" LED display (\ref{fig:trans-display}). We control the display using the Python ST7789 library\footnote{https://github.com/pimoroni/st7789-python}. The beam splitter partially reflects light emitted from the display into the victim camera lens. Light from the perceived object or scene can simultaneously reach the victim camera lens. The gadget is already very compact - slightly larger than an apple watch. Commercial transparent displays for AR can be even smaller (as small as a piece of len in a pair of glasses\footnote{https://www.lx-ar.com/\#/device/1?source\_inside=product}).
% \yi{details of the hardware setup such as splitter.}

\noindent\textbf{Victim Camera.} A Raspberry Pi Camera v2 was selected as the victim camera for all experiments. The camera is controlled by the PiCamera module\footnote{https://picamera.readthedocs.io/en/release-1.13/}. The camera's parameters (e.g., auto exposure, white balance) are fixed for all captured images.

\noindent\textbf{Controlling the Light Conditions.} Indoor experiments were performed in a controlled laboratory environment. We measured the ambient light with a Urceri MT-912 lux meter. By adjusting the scene illumination, we were able to simulate different ambient environments, e.g., sunny versus cloudy conditions. The default environmental luminous flux was approximately $50$ lux. A $65$ Watt floodlight placed at varying distances was used to achieve different levels of illumination, with a maximum illuminance of approximately $3000$ lux. 

\noindent\textbf{}

\subsection{Traffic Sign Recognition Results} 

% In this use case, we assume the victim camera is mounted on an autonomous vehicle. Captured images are fed into a traffic sign recognition module. The goal of \name is to cause misclassifications by the model.

In this experiment, the goal of \name is to cause misclassifications of traffic sign recognition modules.

% \begin{figure}
%     \centering
%     \missingfigure{Confusion matrix of the trained traffic sign recognition model.}
%     \caption{Caption}
%     \label{fig:case1_cm}
% \end{figure}

% \begin{table}
% \centering
%     \begin{tabular}{|c|c|c|c|c|c|c|}
%         \hline
%          & School Zone & Stop Sign & Yield & Slow & ...  \\
%          \hline
%         ASR &  &  &  &  &   \\
%         \hline
%     \end{tabular}
% \caption{ASR of the first use case.}
% \label{table:case1_asr}
% \end{table}

\noindent\textbf{Recognition Modules.} We consider $3$ recognition modules: one classifier and two object detectors. For the classifier, we finetune a Resnet-50~\cite{he2016deep} using the combined dataset of the LISA traffic sign dataset~\cite{mogelmose2012vision} and $3400$ traffic sign images that we collected. The LISA traffic sign dataset is a well-known benchmark for traffic sign recognition applications.  Similar to the LisaCNN~\cite{eykholt2018robust} model, we use $17$ classes from the dataset. The total number of images in all $17$ classes is $6966$. The fine-tuned classifier achieves $99\%$ accuracy, which matches the results of LisaCNN. 
For object detectors, we consider Faster R-CNN~\cite{ren2015faster} and YOLO v3~\cite{redmon2018yolov3}. For Faster R-CNN, we use ResNet-50 and a feature pyramid network~\cite{lin2017feature} as the backbone. For YOLO v3 we use Darknet-53 as the backbone~\cite{redmon2018yolov3}. Similar to~\cite{lovisotto2020slap}, we set the detection threshold of bounding boxes for Faster R-CNN and YOLO v3 to be $0.6$ and $0.4$ respectively. The input image size is set to be $240$x$240$ for Faster R-CNN and ResNet-50, and  $224$x$224$ for YOLO v3.
Both of these two detectors are pretrained on the MS COCO dataset~\cite{lin2014microsoft}. For Faster R-CNN we obtained the pretrained checkpoint from PyTorch~\cite{NEURIPS2019_9015}. For Yolov3, we download the checkpoint from its official GitHub page\footnote{https://github.com/ultralytics/yolov3}. %\eric{How many images from MSCOCO?}\yi{it is not us who trained these detectors, we downloaded pretrained checkpoints} 
Since the MS COCO dataset only contains stop signs but excludes all of the other traffic signs we otherwise consider, for the detectors, we present attack results only for stop signs.

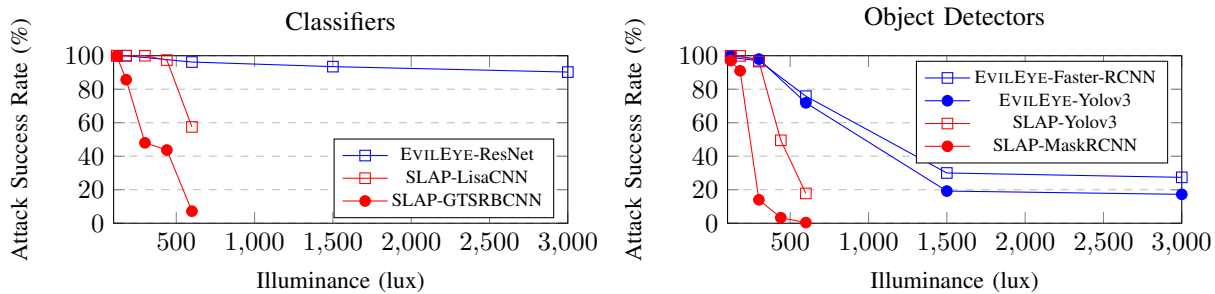
\begin{figure*}[ht]
\begin{tikzpicture}
\begin{axis}[
    label style={font=\small},
    xlabel={Illuminance (lux)},
    ylabel={Attack Success Rate (\%)},
    xmin=100, xmax=3000,
    ymin=0, ymax=100,
    xtick={0, 500, 1000, 1500, 2000, 2500, 3000},
    ytick={0, 20, 40, 60, 80, 100},
    legend pos=south east,
    legend style={nodes={scale=0.7, transform shape}},
    ymajorgrids=true,
    grid style=dashed,
    height=1.5in,
    width=3in,
    title=Classifiers,
    ]

% \addplot[
%     color=blue,
%     mark=square,
%     ]
%     coordinates {
%     (600,99.4)(800,95)(1000,92)(1500,90)(2000,87.5)(3000,70)
%     };

% \addplot[
%     color=red,
%     mark=square,
%     ]
%     coordinates {
%     (120,99)(300,90.6)(600,58.4)
%     };
    
\addplot [color=blue, mark=square] table [x=illuminance, y=asr, col sep=comma] {Data/cat_resnet50.csv};

% \addplot [color=blue, mark=x] table [x=illuminance, y=asr, col sep=comma] {Data/cat_fasterrcnn.csv};

% \addplot [color=blue, mark=+] table [x=illuminance, y=asr, col sep=comma] {Data/cat_yolo.csv};

\addplot [color=red, mark=square] table [x=illuminance, y=asr, col sep=comma] {Data/slap_lisacnn.csv};

\addplot [color=red, mark=*] table [x=illuminance, y=asr, col sep=comma] {Data/slap_gtsrbcnn.csv};

% \addplot [color=red, mark=+] table [x=illuminance, y=asr, col sep=comma] {Data/slap_yolo.csv};

% \addplot [color=red, mark=square] table [x=illuminance, y=asr, col sep=comma] {Data/slap_maskrcnn.csv};

\legend{\namenospace-ResNet, SLAP-LisaCNN, SLAP-GTSRBCNN}
\end{axis}
% \caption{Classifier results.}
\end{tikzpicture}
\begin{tikzpicture}
\begin{axis}[
    label style={font=\small},
    xlabel={Illuminance (lux)},
    ylabel={Attack Success Rate (\%)},
    xmin=100, xmax=3000,
    ymin=0, ymax=100,
    xtick={0, 500, 1000, 1500, 2000, 2500, 3000},
    ytick={0, 20, 40, 60, 80, 100},
    legend pos=north east,
    legend style={nodes={scale=0.7, transform shape}},
    ymajorgrids=true,
    grid style=dashed,
    height=1.5in,
    width=3in,
    title=Object Detectors
    ]
    
\addplot [color=blue, mark=square] table [x=illuminance, y=asr, col sep=comma] {Data/cat_fasterrcnn.csv};

\addplot [color=blue, mark=*] table [x=illuminance, y=asr, col sep=comma] {Data/cat_yolo.csv};

\addplot [color=red, mark=square] table [x=illuminance, y=asr, col sep=comma] {Data/slap_yolo.csv};

\addplot [color=red, mark=*] table [x=illuminance, y=asr, col sep=comma] {Data/slap_maskrcnn.csv};

\legend{\namenospace-Faster-RCNN, \namenospace-Yolov3, SLAP-Yolov3, SLAP-MaskRCNN}
\end{axis}
% \caption{Object Detector results}
\end{tikzpicture}
\centering
\vspace{-0.1in}
\caption{\name indoor attack results of classifiers and object detectors. \name works under various light conditions and can survive significantly stronger ambient light comparing to prior work.}
\vspace{-0.2in}
\label{fig:indoor_asr}
\end{figure*}

\noindent\textbf{Evaluation Protocol.} We created physical models of all $17$ classes of traffic signs that we used to train the recognition module for evaluating our attack. Each model traffic sign is approximately $3$cm x $3$cm. Since an actual traffic sign is much larger (an actual stop sign is about $76$cm x $76$cm), we scaled down the distance between the victim camera and the traffic sign accordingly. For instance, $47$cm in our setup corresponds to $12$m for an actual traffic sign. To collect data for evaluation, we placed a model traffic sign in front of the victim camera as shown in Section~\ref{fig:trans-display}. We recorded videos while moving the traffic sign to different distances and angles. Similar to~\cite{lovisotto2020slap}, we consider a maximum distance of $47$cm (which is equivalent to $12$m for a full-size traffic sign). For varying the angles, we keep the traffic sign within the field of view of the camera. Each video contains on average $500$ frames.
% \eric{How many images/videos were collected?}

\noindent\textbf{Generating Perturbations.} When generating a perturbation using the dot-based perturbation generating function (Section~\ref{sec:modeling}), we set the number of dots $n$ to be 100, the radius of the $k^{th}$ dot $r_k$ to be $1/10$ of the perturbation's height, the alpha blending value $\alpha_{\textnormal{max}}$ to be $1$, and the alpha dropoff $\beta$ to be $1$. We define the attack objective for the traffic sign classifier to minimize the reciprocal of the cross-entropy loss with respect to the true label. For detectors, the attack objective is to minimize the average class score of all the bounding boxes with respect to the class under attack. Our training dataset for generating the perturbations contains $1,000$ images for each type of traffic sign. We use a batch size of 16. The initial learning rate for the dot centers $c$ and color $\gamma$ are $1$ and $0.1$ respectively. Each batch is divided by 10 every $200$ epochs. We train each perturbation for $500$ epochs but we observe the optimization usually converges much earlier.

\noindent\textbf{Indoor Results.} For indoor experiments, we evaluated the adversarial perturbations under various illuminance values (produced by moving the floodlight) ranging from $120$ lux to $3000$ lux. We define the attack success rate (ASR) to be the rate of misclassified frames across all recorded video frames with the target object. %out of all the frames of a video we recorded with a target object in the  tframe, how many of them are misclassified. 
For the traffic sign classifier, this corresponds to the rate of misclassification into any other class. For the traffic sign detector, it corresponds to any missed detections. Figure~\ref{fig:indoor_asr} shows the results of our indoor experiments. For a given model, we report the average ASR over all the traffic signs across different illuminance values. As can be seen from the figures, our attack is effective in a wide range of environmental lighting conditions, including much stronger ambient light when compared to prior work. We attribute this success to our hardware and software co-designed attack approach, which enables a strong perturbation to be computed and precisely imposed on the victim camera. %The attack can survive much stronger ambient light when compared to prior work. 
Because our attack is sensor-based, i.e., the adversarial perturbation is placed in close proximity to the victim camera, the ambient light has less of an impact on the perturbation performance. %Table~\ref{tab:light} to give the readers an idea of what different levels of illuminance mean. 
Although we observe that the ASR for object detectors degrades in strong ambient light, we note that fooling an object detector is more difficult than fooling a classifier since the adversarial perturbation needs to account for both the bounding box and the class of the object in the bounding box. Specifically, only the pixel mutations inside and around the bounding box of the object can effectively alter the final prediction of the model. 

\begin{figure}[htp]
% \begin{minipage}{0.45\textwidth}
\centering
\begin{tikzpicture}
\begin{axis}[
    label style={font=\small},
    xlabel={Illuminance (lux)},
    ylabel={Attack Success Rate (\%)},
    xmin=0, xmax=3000,
    ymin=0, ymax=110,
    xtick={0, 500, 1000, 1500, 2000, 2500, 3000},
    ytick={0, 20, 40, 60, 80, 100},
    legend pos=south west,
    legend style={nodes={scale=0.5, transform shape}},
    ymajorgrids=true,
    grid style=dashed,
    height=1.6in,
    width=3in,
]
    % \draw[solid, color=blue, ultra thick] (0,89.5)--(12,89.5);
    % % \draw[dashed, color=blue] (12,89.5)--(12,0);
    
    % \draw[solid, color=red, ultra thick] (0,77)--(12,77);
    % % \draw[dashed, color=red] (12,77)--(12,0);
    
    % \draw[solid, color=orange, ultra thick] (0,100)--(12,100);
    % % \draw[dashed, color=orange] (12,100)--(12,0);
    
    % \draw[dashed, color=black] (12,110)--(12,0);

    \addplot [color=blue, mark=triangle] table [x=illuminance, y=asr, col sep=comma] {Data/cat_resnet50_outdoor.csv};
    
    \addplot [color=blue, mark=square] table [x=illuminance, y=asr, col sep=comma] {Data/cat_fasterrcnn_outdoor.csv};
    
    \addplot [color=blue, mark=diamond] table [x=illuminance, y=asr, col sep=comma] {Data/cat_yolo_outdoor.csv};

    \addplot [color=red, mark=*] table [x=illuminance, y=asr, col sep=comma] {Data/slap_lisacnn_outdoor.csv};
    
    \addplot [color=red, mark=*] table [x=illuminance, y=asr, col sep=comma] {Data/slap_gtsrbcnn_outdoor.csv};
    
    \addplot [color=red, mark=*] table [x=illuminance, y=asr, col sep=comma] {Data/slap_maskrcnn_outdoor.csv};
    
    \addplot [color=red, mark=*] table [x=illuminance, y=asr, col sep=comma] {Data/slap_maskrcnn_outdoor.csv};
    
    \legend{\namenospace-ResNet, \namenospace-Faster-RCNN, \namenospace-Yolov3, SLAP Models (120 lux)}
    
\end{axis}
\end{tikzpicture}
\vspace{-0.1in}
\caption{\name outdoor attack results. The object classification ASR remains high, while the object detection ASR drops off at higher illuminance values.
}
\vspace{-0.1in}
\label{fig:outdoor_asr}
% \end{minipage} \hfill
\end{figure}
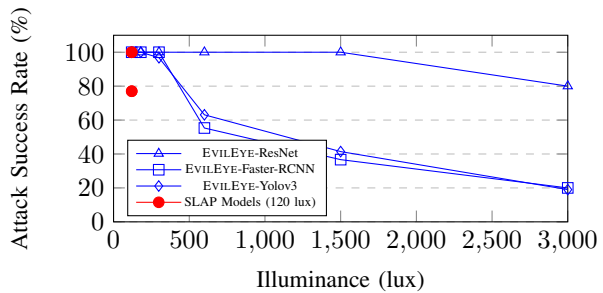

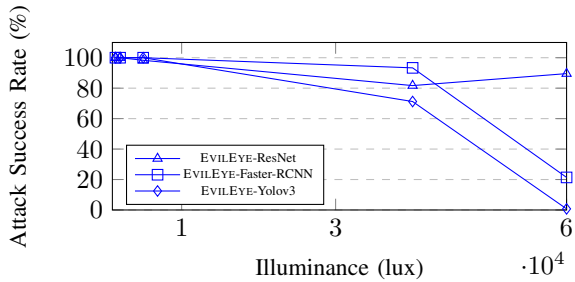
\begin{figure}[t]
\centering
% \begin{minipage}{0.33\textwidth}
\begin{tikzpicture}
\begin{axis}[
    label style={font=\small},
    xlabel={Illuminance (lux)},
    ylabel={Attack Success Rate (\%)},
    xmin=1000, xmax=60000,
    ymin=0, ymax=110,
    xtick={0, 10000, 30000, 60000},
    ytick={0, 20, 40, 60, 80, 100},
    legend pos=south west,
    legend style={nodes={scale=0.5, transform shape}},
    ymajorgrids=true,
    grid style=dashed,
    height=1.5in,
    width=3in,
]

    \addplot [color=blue, mark=triangle] table [x=illuminance, y=asr, col sep=comma] {Data/cat_resnet50_outdoor_p.csv};
    
    \addplot [color=blue, mark=square] table [x=illuminance, y=asr, col sep=comma] {Data/cat_fasterrcnn_outdoor_p.csv};
    
    \addplot [color=blue, mark=diamond] table [x=illuminance, y=asr, col sep=comma] {Data/cat_yolo_outdoor_p.csv};
    
    \legend{\namenospace-ResNet, \namenospace-Faster-RCNN, \namenospace-Yolov3}
    
\end{axis}
\end{tikzpicture}
\vspace{-0.1in}
\caption{Improving the \name outdoor results with the neutral density (ND) filter that attenuates the impact of illuminance. \name can work in direct sunlight ($40$k to $60$k lux).}
\label{fig:filter}
\vspace{-0.15in}
% \end{minipage}\par
\end{figure}

\noindent\textbf{Outdoor Results.} Outdoor experiments allow us to test our attack in a more realistic environment as well as in stronger natural light. To perform outdoor experiments, we transported the entire experimental setup to an outdoor road. We measured the illuminance using the same lux meter. %The outdoor experiments were conducted on a clear sunny day afternoon and evening in May.  
% Since it is not safe and easy for us to collect results on open roads with various traffic signs, we again use our printed signs. It is not possible for us to control the outdoor illuminance precisely, we evenly sample multiple time slots in an afternoon (from 12pm to 6pm) and measure the environmental illuminance during these time slots. 
Figure~\ref{fig:outdoor_asr} shows the outdoor attack results. For the classifier, \name is able to maintain a high ASR across all tested illuminance levels. For detectors, \name has nearly perfect attack performance up to 600 lux. This is equivalent to an overcast afternoon with some overhead cover. As illuminance increases, \name is still effective up to 3,000 lux, which is equivalent to noon on an overcast day. In general, \name can survive much stronger ambient light compared to prior light-based attacks~\cite{lovisotto2020slap}, which were tested to be effective only up to $120$ lux. To push the limit of \name in terms of illuminance, we exposed the setup to direct sunlight. In the most extreme case, we pointed the victim camera towards the sun. The measured illuminance was $60,000$ lux. To reduce the amount of light traveling into the camera lens, we utilized a neutral-density (ND) filter in front of the transparent display. The attack results are shown in Figure~\ref{fig:filter}. From the results with the ND filter, we found that \name can work under direct sunlight.
% \yi{also put clean accuracy as a reference.}
% \eric{unfinished}

\begin{figure}
\begin{tikzpicture}
\begin{axis}[
    label style={font=\small},
    xlabel={Illuminance (lux)},
    ylabel={Attack Success Rate (\%)},
    xmin=0, xmax=3000,
    ymin=60, ymax=110,
    xtick={0, 500, 1000, 1500, 2000, 2500, 3000},
    ytick={60, 80, 100},
    legend pos=south west,
    legend style={nodes={scale=0.5, transform shape}},
    ymajorgrids=true,
    grid style=dashed,
    height=1.5in,
    width=3in,
]

    \addplot [color=blue, mark=square] table [x=illuminance, y=asr, col sep=comma] {Data/dynamic.csv};
    
    \addplot [color=red, mark=+] table [x=illuminance, y=asr, col sep=comma] {Data/static.csv};
    
    \legend{Dynamic, Static}
% \addplot[
%     color=blue,
%     mark=square,
%     ]
%     coordinates {
%     (600,99)(800,99)(1000,95)(1500,85)(2000,80)(3000,70)
%     };
%     \legend{Resnet-50}
    
\end{axis}
\end{tikzpicture}
\centering
\caption{Comparing \name to a static approach. \namenospace's dynamic approach switch perturbations for different type of traffic signs while the static approach crafts one single perturbation to fool all the traffic signs. \namenospace's dynamic approach gives significantly better ASR.}
\label{fig:compare_static}
\vspace{-0.2in}
\end{figure}
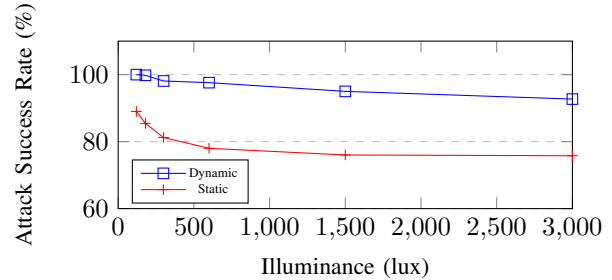

\noindent\textbf{Comparison to Static Approach.} As mentioned in Section~\ref{sec:design}, one major advantage of our attack approach compared to prior work is that our attack is dynamic. The adversary can change the displayed perturbation based on different situations. For example, the adversary can change the perturbations for different traffic signs based on the location of the victim vehicle to maximize the attack damage.
% \eric{what situations? be more specific} 
For static approaches, once a perturbation is attached to the victim camera, the attack cannot be changed or removed. To evaluate the efficacy of our approach, we compared the following two cases: 1) dynamic attack: an adversarial perturbation is crafted for each traffic sign and can be changed dynamically during the attack phase; and 2) static attack: an adversarial perturbation is crafted to maximize the misclassification rate of all types of traffic signs. To prepare such a perturbation, we perform the same optimization process with the same attack objective: maximize the loss with respect to the true label. However, instead of using images from a single class (e.g., the stop sign), we use images from all the classes. We randomly sample 1,000 routes from a traffic sign map\footnote{https://help.mapillary.com/hc/en-us/articles/360003021432-Exploring-traffic-signs}.  Each route contains approximately 100 traffic signs. We then test how effective the perturbations in each case are on these traffic signs. For the dynamic attack, the adversary is able to locate the vehicle through GPS and has access to a traffic sign map. For each encountered traffic sign, the perturbation specifically designed for it can be displayed accordingly. For the static attack, since there is only one single perturbation, it will be used for all the traffic signs. We report the average ASR over all the traffic signs in a route and all the routes for both cases in Figure~\ref{fig:compare_static}. Attempting to use one single perturbation for all the classes results in degraded attack performance because designing a single perturbation to accommodate too many cases is more challenging. \name has the ability to change the perturbation dynamically so each perturbation can focus on its own target class.

\noindent\textbf{Transferability.} We also investigate the transferability of \name across different models--in addition to the $3$ models we already evaluated. We also consider Google Vision\footnote{https://cloud.google.com/vision}, a commercial pre-trained Image content recognizer. %\yi{this sounds like we are only evaluating Google Vision, in fact we are evaluating all 3 models.} 
We queried the Google Vision API with our perturbed images. We define the ASR in this case as the percentage of queried images that do not return ``Stop Sign" in their top predictions. We report the cross-model ASR in Figure~\ref{tab:transferability}. According to the results, perturbations are more transferable between Faster R-CNN and YOLO v3, while not as transferable with ResNet-50. This is because Faster R-CNN and YOLO v3 are both object detectors and are both trained on the same dataset. This makes it easier for AEs to transfer between them. On the contrary, ResNet-50 is performing a different task (i.e., only classification) and trained on a different dataset. Finally, Google Vision is completely unable to defend against \namenospace. 
Results are shown in \autoref{tab:transferability}.

% \usepackage{multirow}
\begin{table}[]
\centering
\caption{Transferability Results. We test generated perturbations across different model architectures. We additionally test transferability on the Google Vision Classifier API, achieving 100\% ASR in all cases. (*) indicates the usage of a neutral-density filter (approx. 90\% light reduction).}
\vspace{-0.1in}
\resizebox{.48\textwidth}{!}{%
\begin{tabular}{cc|cccc}
\hline
\multicolumn{2}{c|}{}                 & \multicolumn{4}{c}{Target Model (ASR)}    \\ \hline
lux                    & Source Model & ResNet-50 & YOLO v3  & Faster R-CNN & Google Vision \\ \hline
\multirow{3}{*}{120}   & ResNet-50    & -         & 68.01\% & 98.32\% & 100\%     \\
                       & YOLO v3       & 0\%       & -       & 100\%  & 100\%     \\
                       & Faster R-CNN & 0\%       & 95.2\%  & -       & 100\%     \\ \hline
\multirow{3}{*}{600}   & ResNet-50    & -         & 13.77\% & 33.33\% & 100\%     \\
                       & YOLO v3       & 0\%       & -       & 86.61\%& 100\%     \\
                       & Faster R-CNN & 0\%       & 32.23\% & -       & 100\%     \\ \hline
\multirow{3}{*}{1400*}  & ResNet-50    & -         & 100\%   & 100\%  & 100\%     \\
                       & YOLO v3       & 0\%       & -       & 100\%  & 100\%     \\
                       & Faster R-CNN & 0\%       & 92.22\% & -       & 100\%     \\ \hline
\multirow{3}{*}{5000*}  & ResNet-50    & -         & 100\%   & 100\%  & 100\%     \\
                       & YOLO v3       & 0\%       & -       & 100\%  & 100\%     \\
                       & Faster R-CNN & 0\%       & 100\%   & -       & 100\%     \\ \hline
\multirow{3}{*}{40000*} & ResNet-50    & -         & 26.77\% & 45.07\%& 100\%     \\
                       & YOLO v3       & 0\%       & -       & 63.31  & 100\%     \\
                       & Faster R-CNN & 3.33\%    & 82.23\% & -       & 100\%     \\ \hline
\end{tabular}
}
\vspace{-0.1in}
\label{tab:transferability}
\end{table}

% \begin{figure}
% \begin{tikzpicture}
% \begin{axis}[
%     % title={Temperature dependence of CuSO\(_4\cdot\)5H\(_2\)O solubility},
%     xlabel={Illuminance},
%     ylabel={Attack Success Rate (ASR)},
%     xmin=500, xmax=3000,
%     ymin=50, ymax=100,
%     xtick={500, 1000, 1500, 2000, 2500, 3000},
%     ytick={50, 60, 70, 80, 90, 100},
%     legend pos=north east,
%     ymajorgrids=true,
%     grid style=dashed,
% ]

% % \addplot[
% %     color=blue,
% %     mark=square,
% %     ]
% %     coordinates {
% %     (600,99)(800,99)(1000,95)(1500,85)(2000,80)(3000,70)
% %     };
% %     \legend{Resnet-50}
    
% \end{axis}
% \end{tikzpicture}
% \caption{Compare with the static approach. (TODO: direct comparisons for reviewers)
% \eric{need to expand caption}
% \yi{3 models, 6 light conditions, each compared with a single perturbation optimized for all traffic signs}}
% \label{fig:compare_static}
% \end{figure}

% \usepackage{multirow}
% \begin{table}[]
% \centering
% \begin{tabular}{lcccccc}
% \hline
% \multirow{2}{*}{}                   & \multicolumn{6}{c}{Illuminance}       \\
%                                     & 120 & 300 & 600 & 1000 & 1500 & 2000 \\ \hline
% \multicolumn{1}{c}{Cataract}        &     &     &     &      &      &      \\
% \multicolumn{1}{c}{Static Approach} &     &     &     &      &      &      \\ \hline
% \end{tabular}
% \caption{Comparing to a static approach.}
% \label{tab:compare_static}
% \end{table}

\begin{table}[]
\footnotesize
\centering
\caption{ASR of our AEs evaluated against existing defense solutions under different light levels. We present averaged runs for Input Randomization, and only the best defensive result for Feature Squeezing (full results can be seen in Figure~\ref{fig:feature-sq}).}
\vspace{-0.1in}
\resizebox{1\linewidth}{!}{%
\begin{tabular}{cc|cccccc}
  \hline
Model &
  \begin{tabular}[c]{@{}c@{}}Ambient\\ Light (lux)\end{tabular} &
  \begin{tabular}[c]{@{}c@{}}Baseline \\ ASR\end{tabular} &
  \begin{tabular}[c]{@{}c@{}}Input \\ Randomization\end{tabular} &
  \begin{tabular}[c]{@{}c@{}}SentiNet \end{tabular} &
  \begin{tabular}[c]{@{}c@{}}Feature \\ Squeezing\end{tabular} &
  %\begin{tabular}[c]{@{}c@{}}Adversarial \\ Training (AT)\end{tabular} &
  \begin{tabular}[c]{@{}c@{}}Physical Adv. \\ Training \end{tabular}
  \\ \hline
\multirow{5}{*}{Resnet-50} & 120  & 100\%  & 98.5\% & 100\%  & 94.1\% &  100\% \\
                           & 180  & 100\%  & 98.3\% & 100\%  & 59.2\% &  100\% \\
                           & 300  & 100\%  & 90.5\% & 100\%  & 84.1\% &  100\% \\
                           & 600  & 100\%  & 95.8\% & 100\%  & 89.9\% & 75.3\% \\
                           & 1500 & 92.7\% & 92.2\% & 87.0\% & 28.3\% & 26.1\% \\
                           %& 3000 & 78.3\% & 80.5\% & 67.5\% & 0\%    & 24.1\% \\
                           & 3000 & 88.7\% & 91.4\% & 67.5\% & 86.8\%    & 34.3\% \\
                           \hline
\end{tabular}
}
\label{tab:defense}
\vspace{-0.1in}
\end{table}

\subsection{Evaluation Against Existing Adversarial ML Defenses}

In this section, we evaluate our physical adversarial perturbations against existing ML defenses. 
% Many defenses focus on digital-domain attacks, where adversaries can precisely perform specific fine-grained modifications (e.g., pixel-level manipulations), whereas few defenses explicitly mention or seek to defend against physical-domain attacks. Therefore, 
We consider two main criteria in choosing defenses to evaluate. The first criteria is non-specificity, as defenses targeting specific attack signatures or features are less likely to be successful against attacks they weren't designed for. The second defense criteria is low computational overhead, allowing it to be used in real-time applications. As such, we evaluate these defenses on traffic sign identification.
Specifically, we test our adversarial perturbations against SentiNet~\cite{chou2020sentinet} -- a defense designed to detect physical AEs (e.g., physically-placed patches or stickers) -- as well as feature-squeezing~\cite{xu2017feature} and input randomization~\cite{xie2017mitigating}, two adversarial defenses which transform classifier inputs and meet the criteria of generality and low overhead. Additionally, we evaluate adversarial training~\cite{goodfellow2014explaining} as a general and preemptive defense simulating a knowledgeable or adaptive defender. A summary of these results can be found in Table~\ref{tab:defense}. In general, the trend of increasing brightness suppresses the effectiveness of \namenospace, resulting in decreased ASR as lux levels increase.

\begin{comment}
\begin{figure}
\centering
\includegraphics[width=1\linewidth]{\figurepath/sentinet_classifier.png}
\caption{A visualization of SentiNet attack detection. Adversarial samples fall within the distribution of benign samples in terms of Average Classifier Confidence and Fool Percentage when salient areas are masked by SentiNet, and are therefore undetected.
%\eric{needs larger font}
}
\label{fig:sentinet-asr}
\end{figure}
\end{comment}

\begin{comment}
\begin{figure}
\hspace*{-12px}
\centering
\begin{subfigure}{.27\textwidth}
  \includegraphics[width=\linewidth]{\figurepath/sentinet_1500.png}
\subcaption{1500 lux}
\end{subfigure}%
\begin{subfigure}{.25\textwidth}
  \includegraphics[width=\linewidth]{\figurepath/sentinet_3000_2.png}
\subcaption{3000 lux}
\end{subfigure}
\caption{A visualization of SentiNet attack detection at different lux levels. Adversarial samples fall within the distribution of benign samples in terms of Average Classifier Confidence and Fool Percentage when salient areas are masked by SentiNet, and are therefore difficult to detect.}
\label{fig:sentinet-asr}
\vspace{-0.15in}
\end{figure}
\end{comment}

\begin{figure}
\centering
  \includegraphics[width=\linewidth]{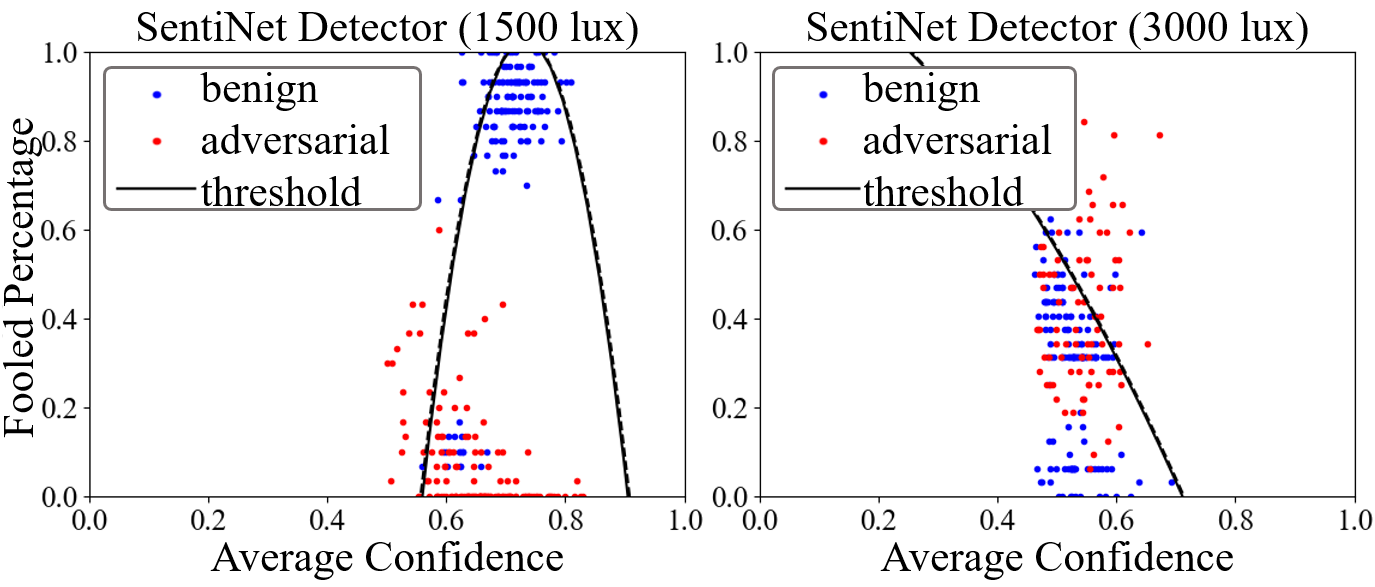}
\caption{A visualization of SentiNet attack detection at different lux levels. Adversarial samples fall within the distribution of benign samples in terms of Average Classifier Confidence and Fool Percentage when salient areas are masked by SentiNet, and are therefore difficult to detect.}
\label{fig:sentinet-asr}
%\vspace{-0.15in}
\end{figure}

\noindent\textbf{SentiNet.}
SentiNet~\cite{lovisotto2020slap} detects adversarial attacks using a comparative saliency mechanism.
Using Classifier Confidence and classifier Fool Rate (\%) to differentiate between benign and adversarial salient regions over several datasets, SentiNet can detect adversarial patches and perturbations.
%Given a small set of known-clean reference images with known labels, an additional set of clean benign images, and a set of adversarial images,
%SentiNet trains a polynomial threshold function over two metrics: Fool Percentage and Classifier Confidence.% Using a saliency function like GradCAM~\cite{selvaraju2017grad} or XRAI~\cite{kapishnikov2019xrai},
%SentiNet isolates salient regions of benign and adversarial image sets, overlaying them onto reference images and noting any changes in network classification (Fool Percentage) and Classifier Confidence. After collecting this data, SentiNet fits a polynomial function separating benign and adversarial sample groups.
%We evaluate SentiNet by first using a set of 200 benign images to fit SentiNet's attack detection classifier. We then evaluate 100 adversarial images using SentiNet's salience region testing. For finer granularity, we compute higher-resolution saliency regions using XRAI, following the implementation of~\cite{lovisotto2020slap}.
Results in Figure \ref{fig:sentinet-asr} show a low detection rate, with the adversarial samples' distribution heavily overlapping and falling below the attack detection threshold polynomial fit by the behavior of the benign samples.% (Fig. 13b). %within the benign distribution for the two discriminatory features of SentiNet.
We suspect that our adversarial perturbations are more difficult to detect using SentiNet because they are less centralized than the original patches studied. This is suggested by the low Fool Percentage of our AEs %(See Appendix B Figure~\ref{fig:feature-sq}) 
-- individually, a single dot or localized dot cluster is not salient enough to produce adversarial behavior. % might be incorrect: Conversely, the slightly higher Average Confidence (evaluated with overlaid(?) salient regions replaced by low-salience Gaussian noise) of our adversarial samples indicate that our perturbations do not function exclusively by occluding the traffic sign. 

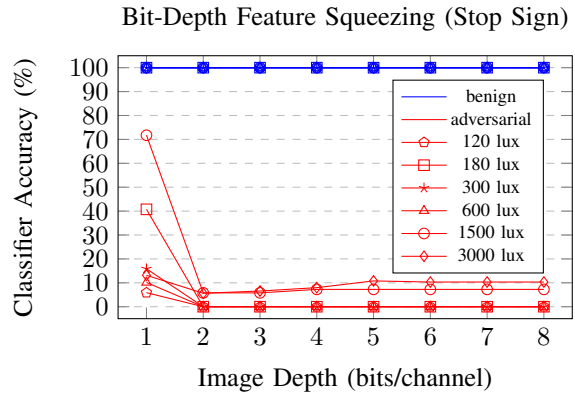
\begin{figure}
\begin{tikzpicture}
\begin{axis}[
    title={Bit-Depth Feature Squeezing (Stop Sign)},
    xlabel={Image Depth (bits/channel)},
    ylabel={Classifier Accuracy (\%)},
    xmin=.5, xmax=8.5,
    ymin=-5, ymax=105,
    xtick={8,7,6,5,4,3,2,1},
    ytick={0, 10,20,30,40,50,60,70,80,90,100},
    %legend pos= north east,
    %legend style={at={(axis cs:8,50)},anchor=north east},
    % legend style={nodes={scale=0.7, transform shape}},
    legend style={at={(axis cs:8,95)},anchor=north east,nodes={scale=0.7, transform shape}},
    ymajorgrids=true,
    grid style=dashed,
    height=2in,
    width=3in,
    ]
% Legend Entries
\addplot[color=blue,]
    coordinates {(-1,-1)};
\addlegendentry{benign}
\addplot[color=red,]
    coordinates {(-1,-1)};
\addlegendentry{adversarial}

%120
\addplot[color=blue, mark=pentagon,forget plot]
    coordinates {(8,100.00)(7,100.00)(6,100.00)(5,100.00)(4,100.00)(3,100.00)(2,100.00)(1,100.00)};
\addplot[color=red, mark=pentagon,]
    coordinates {(8,0.00)(7,0.00)(6,0.00)(5,0.00)(4,0.00)(3,0.00)(2,0.00)(1,5.88)};
\addlegendentry{120 lux}
    
%180
\addplot[color=blue, mark=square,forget plot]
    coordinates {(8,100.00)(7,100.00)(6,100.00)(5,100.00)(4,100.00)(3,100.00)(2,100.00)(1,100.00)};
\addplot[color=red, mark=square,]
    coordinates {(8,0.00)(7,0.00)(6,0.00)(5,0.00)(4,0.00)(3,0.00)(2,0.00)(1,40.74)};
\addlegendentry{180 lux}
    
%300
\addplot[color=blue, mark=star,forget plot]
    coordinates {(8,100.00)(7,100.00)(6,100.00)(5,100.00)(4,100.00)(3,100.00)(2,100.00)(1,100.00)};
\addplot[color=red, mark=star,]
    coordinates {(8,0.00)(7,0.00)(6,0.00)(5,0.00)(4,0.00)(3,0.00)(2,0.00)(1,15.87)};
\addlegendentry{300 lux}

%600
\addplot[color=blue, mark=triangle,forget plot]
    coordinates {(8,100.00)(7,100.00)(6,100.00)(5,100.00)(4,100.00)(3,100.00)(2,100.00)(1,100.00)};
\addplot[color=red, mark=triangle,]
    coordinates {(8,0.00)(7,0.00)(6,0.00)(5,0.00)(4,0.00)(3,0.00)(2,0.00)(1,10.14)};
\addlegendentry{600 lux}

%1500
\addplot[color=blue, mark=o,forget plot]
    coordinates {(8,100.00)(7,100.00)(6,100.00)(5,100.00)(4,100.00)(3,100.00)(2,100.00)(1,100.00)};
\addplot[color=red, mark=o,]
    coordinates {(8,7.25)(7,7.25)(6,7.25)(5,7.25)(4,7.25)(3,5.80)(2,5.80)(1,71.74)};
\addlegendentry{1500 lux}

%3000
\addplot[color=blue, mark=diamond,forget plot]
    coordinates {(8,100.00)(7,100.00)(6,100.00)(5,100.00)(4,100.00)(3,100.00)(2,100.00)(1,100.00)};
\addplot[color=red, mark=diamond,]
    %coordinates {(8,21.69)(7,21.69)(6,21.69)(5,21.69)(4,20.48)(3,20.48)(2,22.89)(1,100.00)};
    coordinates {(8,10.33)(7,10.33)(6,10.33)(5,10.80)(4,7.98)(3,6.57)(2,5.63)(1,13.15)};
\addlegendentry{3000 lux}

%\legend{Benign, Adversarial}
    
\end{axis}
\end{tikzpicture}
\caption{A comparison of classifier accuracy for benign and adversarial samples when varying image bit-depth of the RGB channels, ranging from native 8-bit color resolution (256 possible values/channel) to 1-bit (2 possible values/channel). Color represents benign (blue) and adversarial (red) performance. Markers indicate different lux levels. 
%As bit-depth decreases, so does benign classifier accuracy. However, the classifier consistently performs poorly on adversarially perturbed inputs.
%\eric{can we include a red/blue key in the figure?}
}
\vspace{-0.1in}
\label{fig:feature-sq}
\end{figure}

\noindent\textbf{Feature-Squeezing.} Feature squeezing reduces the dimensionality of classifier inputs, restricting the space for AEs. This technique offers a simple and general strategy to attenuate both physical-domain and digital-domain attacks. We evaluate the effectiveness of bit-depth reduction on our adversarial perturbations. It can be seen that feature squeezing works only at low bit-depth. %We have some more detailed analysis on this defense solution in Appendix B.
% by comparing its impact on classifier accuracy for benign and adversarial samples. 
% Results are shown in Figure~\ref{fig:feature-sq}. 
%As the input bit-depth decreases, the overall classifier accuracy is increasingly affected. This has a slight effect on adversarial samples, however classifier accuracy remains low.

% As the input image bit depth decreases, both benign and adversarial classifier accuracy remains relatively stable until the bit depth drops below 2 bits/channel. At this point, we observe a general increase in adversarial accuracy.
% We hypothesize that this is due to the dot perturbation's color robustness -- each dot maintains a constant color gradually fading outwards. As a result, decreasing the bit-depth only slightly attenuates the perturbation until the final step, in which each channel only has 2 possible values, more significantly affecting the perturbation.
%Notably, feature-squeezing has a moderate effect even at high bit-depths and greater effectiveness at very low bit-depth. However, the steep accuracy-detection tradeoff reduces the practicality of low bit-depth.  %However, our adversarial perturbations robust enough that they are not attenuated until the bit-depth (and classifier accuracy) have been significantly reduced.

\noindent\textbf{Input Randomization.} Input randomization modifies the classifier input during inference with a random re-scaling and padding in order to disrupt possible attacks which may rely on precise positioning. We rescale input images randomly between 224-240 pixels from their original size of 224, then randomly pad each edge to reach the final size of 240. We present averaged results over several runs with different random initializations, however, we note that the variance was less than 10\% among each group. We find that input randomization leads to only minor changes in the effectiveness of our adversarial perturbations.
%We find that input randomization performs well, successfully increasing adversarial classification accuracy to 73.9\% (averaged over multiple runs).
%However, this comes at the expense of clean classifier accuracy, which on average decreases 10\% from the baseline.

\noindent\textbf{Adversarial Training.} 
%Adversarial training incorporates adversarial examples into the training process in order to make classifiers more robust. However, this usually comes at the expense of clean accuracy and is tuned for digital-domain attacks.
Wu \etal\cite{Wu2020Defending} propose an occlusion-based adversarial training process to increase classifier robustness against physically-realizable attacks. 
This method of \textit{physical} adversarial training was designed to combat patch-based and sticker-based attacks~\cite{sharif2016accessorize, eykholt2018robust} with the idea that the characteristics of digital and physical attacks differ greatly, and therefore conventional adversarial training is not well-suited to physical attacks. We find physical adversarial training to be a promising method for partially mitigating our attack, especially at higher lux levels. We hypothesize that the higher effectiveness is hindered by the slight mismatch in threat models, as physical adversarial training focuses on ``occlusive" perturbations like stickers and patches, whereas light-based perturbations are additive and do not occlude.

% \subsection{Trade off between accuracy and speed}

\section{Related Work}\label{sec:related}

%\subsection{Adversarial Example Generation}
%\noindent\textbf{Digital Adversarial Examples.} 
Since the concept of AEs was initially proposed~\cite{szegedy2013intriguing}, many techniques have been put forward to generate AEs against state-of-the-art machine learning classifiers in the digital space~\cite{wang2019security}, where attackers have precise control over adversarial modifications. In this paper, we focus on AEs in the physical domain.

%~\cite{goodfellow2014explaining, madry2017towards}. % FGSM, BIM, ILLCM, PGD, C&W
% Move next line to intro or bg
%Designed for attacking in the digital space, the resulting adversarial examples are fed directly to the target classifier and can be precisely controlled by the attacker, even down to the modification of a single pixel~\cite{su2019one}.

%\begin{itemize}
%    \item linear properties of classifier manifolds
%\end{itemize}

\noindent\textbf{Physical Adversarial Examples.}
% In this section, we survey existing efforts of transferring adversarial examples to the physical world. 
Kurakin et al.~\cite{kurakin2018adversarial} study how digital-domain attacks perform in the physical world using photographs of printed-out AEs, finding that they can survive certain transformations like sensor noise, rescaling, and brightness changes.
However, Lu et al.~\cite{lu2017no} showed that the robustness of adversarial examples generated using this approach is decreased over multiple viewings from different angles and distances. In order to generate more robust AEs in the real world, Athalye et al.~\cite{athalye2018synthesizing} propose Expectation over Transformation, optimizing AEs over a set of transformations to account for varying physical environmental conditions by projecting 2-dimensional perturbations onto 3D-printed objects.~\cite{jia2022fooling} also, consider a set of environmental transformations when fabricating malicious traffic signs. In all cases, these approaches focused on static, object-modifying attacks.

Many works create physical AEs by applying adversarial patterns directly to objects themselves. For example, Sharif et al.~\cite{sharif2016accessorize} print adversarial patterns on eyeglass frames to fool facial recognition, and Eykholt et al.~\cite{eykholt2018robust} place stickers on street signs mimicking graffiti to cause misclassification. %Chen et al.~\cite{chen2018shapeshifter}
Another set of approaches~\cite{thys2019fooling, brown2017adversarial,zhao2019seeing} generate adversarial patches. Li et al.~\cite{li2019adversarial} demonstrate an adversarial sticker not limited to a single object by applying a translucent sticker in front of the camera itself. Similarly, a key constraint of all these attacks is that they are object-modifying, static attacks, and cannot adapt or be modified once applied.
%Athalye et al~\cite{athalye2018synthesizing} 3D-print objects with adversarial patterns on the surface.

Other works use light-based perturbations projected onto surfaces in facial recognition  attacks~\cite{nguyen2020adversarial,zhou2018invisible} 
and object detection and classification~\cite{lovisotto2020slap, nichols2018projecting}. Projecting AEs onto objects has the benefits of being more dynamic and not leaving behind physical artifacts. However, they suffer from increased sensitivity to environmental lighting conditions and are still difficult to scale to multiple objects. The idea of using light projection to attack has also been applied to the camera sensors. Exploiting ghost effect and auto-exposure control in optical imaging systems, Man et al.~\cite{man2020ghostimage} use a projector to inject arbitrary patterns (e.g. a stop sign) into the victim camera's field of view. This attack suffers from increased sensitivity to environmental lighting conditions as well. Also, it is not easy to apply the attack in a real-world setting (e.g. a moving camera on an autonomous vehicle).  Wang et al.~\cite{wang2021can} utilized infrared (IR) lights to attack autonomous vehicles. However, IR lights can be strongly interfered by solar radiation. This significantly limits the capability of the attack. Along the direction of attacking the vulnerabilities in sensors, Kohler et al.~\cite{kohler2021they, sayles2021invisible} exploited the rolling shutter in CMOS image sensors using a bright, modulated light source to cause image disruptions. Ji et al.~\cite{ji2021poltergeist} exploited the inertial sensors  meant for image stabilization. In their attack, an adversary controls the output of an inertial sensor by emitting deliberately designed acoustic signals. This produces blurred images that will be misclassified in the decision-making pipeline.

\section{Limitations and Future Work}\label{sec:discussion}
In this section, we discuss the limitations of \name and enumerate future research directions toward bridging the gap between digital domain- and physical domain-based adversarial machine learning.

\noindent\textbf{Practicality of transparent display implementation.} The prototype implementation of \namenospace's transparent provided a low-cost and portable implementation to enable proof-of-concept research for sensor-level, dynamic adversarial attacks. Emerging technologies such as the Microsoft HoloLens~\cite{noor2016hololens} or the Google Glass~\cite{muensterer2014google} have shown the industry trend to develop compact and computationally efficient implementations of transparent displays. Future work could also incorporate advancements in attenuating the impact of ambient light on perception, especially in the context of outdoor displays~\cite{lanca2019effects,eun2016bright}. Thus, the emergence of such technologies would enable the adversaries to implement \name in real-world settings at scale. Also, Itoh et. al \cite{itoh2021towards} provide a comprehensive overview of optical see-through head-mounted displays, comparing designs that tradeoff between various factors such as form factor, size, and light efficiency. Future work can explore a more optimal design across these tradeoffs.

\noindent\textbf{Attack imperceptibility for autonomous systems.} A common notion in adversarial examples research is to ensure perturbations are \textit{imperceptible} to humans in the perception loop. For instance, adversarial examples for spam detectors aimed to bypass machine learning classifiers while not looking suspicious to the target human~\cite{tygar2011adversarial}. For autonomous systems, humans are not expected to be in the loop, i.e., they are not expected to monitor the video camera feed that is being fed into the perception pipeline. Even if a human was monitoring the video feed, the perturbations and objects would most likely be fleeting. Thus, imperceptibility in the context of cyber-physical autonomous systems may target humans who are performing post-incident analyses to investigate likely causes for a malfunction. However, certain autonomous systems domains and applications still require imperceptibility of perturbations, e.g., automatic speech recognition attacks on smart home assistants should be imperceptible to humans who are in the same room~\cite{qin2019imperceptible}. Future work can investigate the need for perceptibility across autonomous systems to understand to \textit{whom} and \textit{when} perturbations need to be imperceptible.   

\noindent\textbf{Real-world considerations.} Several challenges arise when applying \name to the real-world. First, the impact of the additional display on lighting, real-world impact, such as additional reflections need to be considered. \name employs the digital-to-physical mapping to deal with these factors. Moreover, the data used for generating the perturbations and the actual video feed frames at runtime can be different, known as the distributional drift problem. One solution to this can be obtaining more comprehensive training dataset. Also, in some scenarios accessing the victim camera can be challenging for \name. But it remains a big threat to those systems where camera sensors are accessible (e.g. surveillance, authentication, etc.). Also setting changes on the victim system, such as hardware revisions, and software updates can degrade the accuracy of \name. In this case, iterations of the attack will be needed. Currently, \name generates perturbations that can only work on a specific camera. it can be beneficial to make the attack transferable across different cameras in future work. Finally, more complex perception systems employ more complicated sensor arrays (e.g., multi-camera, multi-modal) as the perceptual module. Future work can explore camera pipeline attacks on such systems. 

\section{Conclusion}
This paper introduced \namenospace, a man-in-the-middle perception attack on safety-critical cyber-physical systems. \name is the first sensor-first, dynamic adversarial machine learning framework for physical-domain attacks. \name leverages transparent displays to generate dynamic physical adversarial examples. The digital-to-physical perturbation pipeline is enabled by modeling the environmental noise due to optical transformations and environmental factors. We show the efficacy of \name on the real-world use case of traffic sign recognition, demonstrating that
% \name can dynamically determine which perturbations to generate at runtime without predetermined knowledge of which objects the victim system will encounter. Moreover, we demonstrate that 
\name can significantly outperform existing attack frameworks across varying levels of ambient illumination, including over 80\% ASR at 60,000 lux--whereas prior works can only achieve a similar ASR at less than 120 lux. \name provides a practical approach to dynamically modeling optical transformations in the context of adversarial machine learning attacks in the real world.%\nils{can we bring in some concrete performance numbers/ lux numbers here that highlight some hard evidence of success?}
\section{Acknowledgement}
The research reported in this paper was sponsored in part by the National Science Foundation (NSF) under Award \#2124252 and NSF Secure and Trustworthy Cyberspace (SaTC) program awards, such as \#1705135,; as well as the IoBT REIGN Collaborative Research Alliance funded by the Army Research Laboratory (ARL) under Cooperative Agreement W911NF1720196. The views and conclusions contained in this document are those of the authors and should not be interpreted as representing the official policies, either expressed or implied, of NSF or the U.S. Government. The U.S. Government is authorized to reproduce and distribute reprints for Government purposes notwithstanding any copyright notation here on.

\bibliographystyle{IEEEtrans}
% \bibliographystyle{unsrt}
% \setcitestyle{numbers}
\bibliography{Sources/references.bib}

% Generated by IEEEtranS.bst, version: 1.12 (2007/01/11)
\begin{thebibliography}{10}
\providecommand{\url}[1]{#1}
\csname url@samestyle\endcsname
\providecommand{\newblock}{\relax}
\providecommand{\bibinfo}[2]{#2}
\providecommand{\BIBentrySTDinterwordspacing}{\spaceskip=0pt\relax}
\providecommand{\BIBentryALTinterwordstretchfactor}{4}
\providecommand{\BIBentryALTinterwordspacing}{\spaceskip=\fontdimen2\font plus
\BIBentryALTinterwordstretchfactor\fontdimen3\font minus
  \fontdimen4\font\relax}
\providecommand{\BIBforeignlanguage}[2]{{%
\expandafter\ifx\csname l@#1\endcsname\relax
\typeout{** WARNING: IEEEtranS.bst: No hyphenation pattern has been}%
\typeout{** loaded for the language `#1'. Using the pattern for}%
\typeout{** the default language instead.}%
\else
\language=\csname l@#1\endcsname
\fi
#2}}
\providecommand{\BIBdecl}{\relax}
\BIBdecl

\bibitem{abdi2015vehicle}
L.~Abdi, F.~B. Abdallah, and A.~Meddeb, ``In-vehicle augmented reality traffic
  information system: a new type of communication between driver and vehicle,''
  \emph{Procedia Computer Science}, vol.~73, pp. 242--249, 2015.

\bibitem{athalye2018synthesizing}
A.~Athalye, L.~Engstrom, A.~Ilyas, and K.~Kwok, ``Synthesizing robust
  adversarial examples,'' in \emph{International conference on machine
  learning}.\hskip 1em plus 0.5em minus 0.4em\relax PMLR, 2018, pp. 284--293.

\bibitem{bagloee2016autonomous}
S.~A. Bagloee, M.~Tavana, M.~Asadi, and T.~Oliver, ``Autonomous vehicles:
  challenges, opportunities, and future implications for transportation
  policies,'' \emph{Journal of modern transportation}, vol.~24, no.~4, pp.
  284--303, 2016.

\bibitem{banks2018driver}
V.~A. Banks, K.~L. Plant, and N.~A. Stanton, ``Driver error or designer error:
  Using the perceptual cycle model to explore the circumstances surrounding the
  fatal tesla crash on 7th may 2016,'' \emph{Safety science}, vol. 108, pp.
  278--285, 2018.

\bibitem{brown2017adversarial}
T.~B. Brown, D.~Man{\'e}, A.~Roy, M.~Abadi, and J.~Gilmer, ``Adversarial
  patch,'' \emph{arXiv preprint arXiv:1712.09665}, 2017.

\bibitem{chen2018shapeshifter}
S.-T. Chen, C.~Cornelius, J.~Martin, and D.~H.~P. Chau, ``Shapeshifter: Robust
  physical adversarial attack on faster r-cnn object detector,'' in \emph{Joint
  European Conference on Machine Learning and Knowledge Discovery in
  Databases}.\hskip 1em plus 0.5em minus 0.4em\relax Springer, 2018, pp.
  52--68.

\bibitem{chou2020sentinet}
E.~Chou, F.~Tramer, and G.~Pellegrino, ``Sentinet: Detecting localized
  universal attacks against deep learning systems,'' in \emph{2020 IEEE
  Security and Privacy Workshops (SPW)}.\hskip 1em plus 0.5em minus 0.4em\relax
  IEEE, 2020, pp. 48--54.

\bibitem{deng2020analysis}
Y.~Deng, X.~Zheng, T.~Zhang, C.~Chen, G.~Lou, and M.~Kim, ``An analysis of
  adversarial attacks and defenses on autonomous driving models,'' in
  \emph{2020 IEEE International Conference on Pervasive Computing and
  Communications (PerCom)}.\hskip 1em plus 0.5em minus 0.4em\relax IEEE, 2020,
  pp. 1--10.

\bibitem{ertler2020mapillary}
C.~Ertler, J.~Mislej, T.~Ollmann, L.~Porzi, G.~Neuhold, and Y.~Kuang, ``The
  mapillary traffic sign dataset for detection and classification on a global
  scale,'' in \emph{European Conference on Computer Vision}.\hskip 1em plus
  0.5em minus 0.4em\relax Springer, 2020, pp. 68--84.

\bibitem{eun2016bright}
L.~H. Eun and K.~J. Hong, ``How bright of luminance is needed for outdoor
  commercial display?'' in \emph{2016 IEEE 6th International Conference on
  Consumer Electronics-Berlin (ICCE-Berlin)}.\hskip 1em plus 0.5em minus
  0.4em\relax IEEE, 2016, pp. 141--144.

\bibitem{eykholt2018robust}
K.~Eykholt, I.~Evtimov, E.~Fernandes, B.~Li, A.~Rahmati, C.~Xiao, A.~Prakash,
  T.~Kohno, and D.~Song, ``Robust physical-world attacks on deep learning
  visual classification,'' in \emph{Proceedings of the IEEE Conference on
  Computer Vision and Pattern Recognition}, 2018, pp. 1625--1634.

\bibitem{goodfellow2016deep}
I.~Goodfellow, Y.~Bengio, and A.~Courville, \emph{Deep learning}.\hskip 1em
  plus 0.5em minus 0.4em\relax MIT press, 2016.

\bibitem{goodfellow2014explaining}
I.~J. Goodfellow, J.~Shlens, and C.~Szegedy, ``Explaining and harnessing
  adversarial examples,'' \emph{arXiv preprint arXiv:1412.6572}, 2014.

\bibitem{he2016deep}
K.~He, X.~Zhang, S.~Ren, and J.~Sun, ``Deep residual learning for image
  recognition,'' in \emph{Proceedings of the IEEE conference on computer vision
  and pattern recognition}, 2016, pp. 770--778.

\bibitem{hoory2020dynamic}
S.~Hoory, T.~Shapira, A.~Shabtai, and Y.~Elovici, ``Dynamic adversarial patch
  for evading object detection models,'' \emph{arXiv preprint
  arXiv:2010.13070}, 2020.

\bibitem{hore2010image}
A.~Hore and D.~Ziou, ``Image quality metrics: Psnr vs. ssim,'' in \emph{2010
  20th international conference on pattern recognition}.\hskip 1em plus 0.5em
  minus 0.4em\relax IEEE, 2010, pp. 2366--2369.

\bibitem{huang2011adversarial}
L.~Huang, A.~D. Joseph, B.~Nelson, B.~I. Rubinstein, and J.~D. Tygar,
  ``Adversarial machine learning,'' in \emph{Proceedings of the 4th ACM
  workshop on Security and artificial intelligence}, 2011, pp. 43--58.

\bibitem{ignatov2020rendering}
A.~Ignatov, J.~Patel, and R.~Timofte, ``Rendering natural camera bokeh effect
  with deep learning,'' in \emph{Proceedings of the IEEE/CVF Conference on
  Computer Vision and Pattern Recognition Workshops}, 2020, pp. 418--419.

\bibitem{itoh2021towards}
Y.~Itoh, T.~Langlotz, J.~Sutton, and A.~Plopski, ``Towards indistinguishable
  augmented reality: A survey on optical see-through head-mounted displays,''
  \emph{ACM Computing Surveys (CSUR)}, vol.~54, no.~6, pp. 1--36, 2021.

\bibitem{ji2021poltergeist}
X.~Ji, Y.~Cheng, Y.~Zhang, K.~Wang, C.~Yan, W.~Xu, and K.~Fu, ``Poltergeist:
  Acoustic adversarial machine learning against cameras and computer vision,''
  in \emph{2021 IEEE Symposium on Security and Privacy (SP)}.\hskip 1em plus
  0.5em minus 0.4em\relax IEEE, 2021, pp. 160--175.

\bibitem{jia2022fooling}
W.~Jia, Z.~Lu, H.~Zhang, Z.~Liu, J.~Wang, and G.~Qu, ``Fooling the eyes of
  autonomous vehicles: Robust physical adversarial examples against traffic
  sign recognition systems,'' \emph{arXiv preprint arXiv:2201.06192}, 2022.

\bibitem{kingma2014adam}
D.~P. Kingma and J.~Ba, ``Adam: A method for stochastic optimization,''
  \emph{arXiv preprint arXiv:1412.6980}, 2014.

\bibitem{kohler2021they}
S.~K{\"o}hler, G.~Lovisotto, S.~Birnbach, R.~Baker, and I.~Martinovic, ``They
  see me rollin’: Inherent vulnerability of the rolling shutter in cmos image
  sensors,'' in \emph{Annual Computer Security Applications Conference}, 2021,
  pp. 399--413.

\bibitem{kurakin2018adversarial}
A.~Kurakin, I.~J. Goodfellow, and S.~Bengio, ``Adversarial examples in the
  physical world,'' in \emph{Artificial intelligence safety and
  security}.\hskip 1em plus 0.5em minus 0.4em\relax Chapman and Hall/CRC, 2018,
  pp. 99--112.

\bibitem{lanca2019effects}
C.~Lanca, A.~Teo, A.~Vivagandan, H.~M. Htoon, R.~P. Najjar, D.~P. Spiegel,
  S.-H. Pu, and S.-M. Saw, ``The effects of different outdoor environments,
  sunglasses and hats on light levels: Implications for myopia prevention,''
  \emph{Translational vision science \& technology}, vol.~8, no.~4, pp. 7--7,
  2019.

\bibitem{li2019adversarial}
J.~Li, F.~Schmidt, and Z.~Kolter, ``Adversarial camera stickers: A physical
  camera-based attack on deep learning systems,'' in \emph{International
  Conference on Machine Learning}, 2019, pp. 3896--3904.

\bibitem{li2018critical}
X.~Li, W.~Yi, H.-L. Chi, X.~Wang, and A.~P. Chan, ``A critical review of
  virtual and augmented reality (vr/ar) applications in construction safety,''
  \emph{Automation in Construction}, vol.~86, pp. 150--162, 2018.

\bibitem{lin2017feature}
T.-Y. Lin, P.~Doll{\'a}r, R.~Girshick, K.~He, B.~Hariharan, and S.~Belongie,
  ``Feature pyramid networks for object detection,'' in \emph{Proceedings of
  the IEEE conference on computer vision and pattern recognition}, 2017, pp.
  2117--2125.

\bibitem{lin2014microsoft}
T.-Y. Lin, M.~Maire, S.~Belongie, J.~Hays, P.~Perona, D.~Ramanan,
  P.~Doll{\'a}r, and C.~L. Zitnick, ``Microsoft coco: Common objects in
  context,'' in \emph{European conference on computer vision}.\hskip 1em plus
  0.5em minus 0.4em\relax Springer, 2014, pp. 740--755.

\bibitem{lovisotto2020slap}
G.~Lovisotto, H.~Turner, I.~Sluganovic, M.~Strohmeier, and I.~Martinovic,
  ``Slap: Improving physical adversarial examples with short-lived adversarial
  perturbations,'' \emph{arXiv preprint arXiv:2007.04137}, 2020.

\bibitem{lu2017no}
J.~Lu, H.~Sibai, E.~Fabry, and D.~Forsyth, ``No need to worry about adversarial
  examples in object detection in autonomous vehicles,'' \emph{arXiv preprint
  arXiv:1707.03501}, 2017.

\bibitem{man2020ghostimage}
Y.~Man, M.~Li, and R.~Gerdes, ``$\{$GhostImage$\}$: Remote perception attacks
  against camera-based image classification systems,'' in \emph{23rd
  International Symposium on Research in Attacks, Intrusions and Defenses (RAID
  2020)}, 2020, pp. 317--332.

\bibitem{mehdipour2016comprehensive}
M.~Mehdipour~Ghazi and H.~Kemal~Ekenel, ``A comprehensive analysis of deep
  learning based representation for face recognition,'' in \emph{Proceedings of
  the IEEE conference on computer vision and pattern recognition workshops},
  2016, pp. 34--41.

\bibitem{mogelmose2012vision}
A.~Mogelmose, M.~M. Trivedi, and T.~B. Moeslund, ``Vision-based traffic sign
  detection and analysis for intelligent driver assistance systems:
  Perspectives and survey,'' \emph{IEEE Transactions on Intelligent
  Transportation Systems}, vol.~13, no.~4, pp. 1484--1497, 2012.

\bibitem{moosavi2017universal}
S.-M. Moosavi-Dezfooli, A.~Fawzi, O.~Fawzi, and P.~Frossard, ``Universal
  adversarial perturbations,'' in \emph{Proceedings of the IEEE conference on
  computer vision and pattern recognition}, 2017, pp. 1765--1773.

\bibitem{muensterer2014google}
O.~J. Muensterer, M.~Lacher, C.~Zoeller, M.~Bronstein, and J.~K{\"u}bler,
  ``Google glass in pediatric surgery: an exploratory study,''
  \emph{International journal of surgery}, vol.~12, no.~4, pp. 281--289, 2014.

\bibitem{nassi2020phantom}
B.~Nassi, D.~Nassi, R.~Ben-Netanel, Y.~Mirsky, O.~Drokin, and Y.~Elovici,
  ``Phantom of the adas: Phantom attacks on driver-assistance systems.''
  \emph{IACR Cryptol. ePrint Arch.}, vol. 2020, p.~85, 2020.

\bibitem{nguyen2020adversarial}
D.-L. Nguyen, S.~S. Arora, Y.~Wu, and H.~Yang, ``Adversarial light projection
  attacks on face recognition systems: A feasibility study,'' in
  \emph{Proceedings of the IEEE/CVF Conference on Computer Vision and Pattern
  Recognition Workshops}, 2020, pp. 814--815.

\bibitem{nichols2018projecting}
N.~Nichols and R.~Jasper, ``Projecting trouble: Light based adversarial attacks
  on deep learning classifiers,'' \emph{arXiv preprint arXiv:1810.10337}, 2018.

\bibitem{noor2016hololens}
A.~K. Noor, ``The hololens revolution,'' \emph{Mechanical Engineering}, vol.
  138, no.~10, pp. 30--35, 2016.

\bibitem{NEURIPS2019_9015}
\BIBentryALTinterwordspacing
A.~Paszke, S.~Gross, F.~Massa, A.~Lerer, J.~Bradbury, G.~Chanan, T.~Killeen,
  Z.~Lin, N.~Gimelshein, L.~Antiga, A.~Desmaison, A.~Kopf, E.~Yang, Z.~DeVito,
  M.~Raison, A.~Tejani, S.~Chilamkurthy, B.~Steiner, L.~Fang, J.~Bai, and
  S.~Chintala, ``Pytorch: An imperative style, high-performance deep learning
  library,'' in \emph{Advances in Neural Information Processing Systems 32},
  H.~Wallach, H.~Larochelle, A.~Beygelzimer, F.~d\textquotesingle
  Alch\'{e}-Buc, E.~Fox, and R.~Garnett, Eds.\hskip 1em plus 0.5em minus
  0.4em\relax Curran Associates, Inc., 2019, pp. 8024--8035. [Online].
  Available:
  \url{http://papers.neurips.cc/paper/9015-pytorch-an-imperative-style-high-performance-deep-learning-library.pdf}
\BIBentrySTDinterwordspacing

\bibitem{pendleton2017perception}
S.~D. Pendleton, H.~Andersen, X.~Du, X.~Shen, M.~Meghjani, Y.~H. Eng, D.~Rus,
  and M.~H. Ang, ``Perception, planning, control, and coordination for
  autonomous vehicles,'' \emph{Machines}, vol.~5, no.~1, p.~6, 2017.

\bibitem{qin2019imperceptible}
Y.~Qin, N.~Carlini, G.~Cottrell, I.~Goodfellow, and C.~Raffel, ``Imperceptible,
  robust, and targeted adversarial examples for automatic speech recognition,''
  in \emph{International conference on machine learning}.\hskip 1em plus 0.5em
  minus 0.4em\relax PMLR, 2019, pp. 5231--5240.

\bibitem{redmon2018yolov3}
J.~Redmon and A.~Farhadi, ``Yolov3: An incremental improvement,'' \emph{arXiv
  preprint arXiv:1804.02767}, 2018.

\bibitem{ren2015faster}
S.~Ren, K.~He, R.~Girshick, and J.~Sun, ``Faster r-cnn: Towards real-time
  object detection with region proposal networks,'' \emph{Advances in neural
  information processing systems}, vol.~28, pp. 91--99, 2015.

\bibitem{ronneberger2015u}
O.~Ronneberger, P.~Fischer, and T.~Brox, ``U-net: Convolutional networks for
  biomedical image segmentation,'' in \emph{International Conference on Medical
  image computing and computer-assisted intervention}.\hskip 1em plus 0.5em
  minus 0.4em\relax Springer, 2015, pp. 234--241.

\bibitem{sayles2021invisible}
A.~Sayles, A.~Hooda, M.~Gupta, R.~Chatterjee, and E.~Fernandes, ``Invisible
  perturbations: Physical adversarial examples exploiting the rolling shutter
  effect,'' in \emph{Proceedings of the IEEE/CVF Conference on Computer Vision
  and Pattern Recognition}, 2021, pp. 14\,666--14\,675.

\bibitem{sharif2016accessorize}
M.~Sharif, S.~Bhagavatula, L.~Bauer, and M.~K. Reiter, ``Accessorize to a
  crime: Real and stealthy attacks on state-of-the-art face recognition,'' in
  \emph{Proceedings of the 2016 acm sigsac conference on computer and
  communications security}, 2016, pp. 1528--1540.

\bibitem{souman2021human}
J.~Souman, M.~van Weperen, J.~Hogema, M.~Hoedemaeker, F.~Westerhuis,
  A.~Stuiver, and D.~de~Waard, ``Human factors guidelines report 2: Driver
  support systems overview,'' 2021.

\bibitem{sreenu2019intelligent}
G.~Sreenu and M.~S. Durai, ``Intelligent video surveillance: a review through
  deep learning techniques for crowd analysis,'' \emph{Journal of Big Data},
  vol.~6, no.~1, pp. 1--27, 2019.

\bibitem{szegedy2013intriguing}
C.~Szegedy, W.~Zaremba, I.~Sutskever, J.~Bruna, D.~Erhan, I.~Goodfellow, and
  R.~Fergus, ``Intriguing properties of neural networks,'' \emph{arXiv preprint
  arXiv:1312.6199}, 2013.

\bibitem{thomas2012survey}
B.~H. Thomas, ``A survey of visual, mixed, and augmented reality gaming,''
  \emph{Computers in Entertainment (CIE)}, vol.~10, no.~1, pp. 1--33, 2012.

\bibitem{thys2019fooling}
S.~Thys, W.~Van~Ranst, and T.~Goedem{\'e}, ``Fooling automated surveillance
  cameras: adversarial patches to attack person detection,'' in
  \emph{Proceedings of the IEEE Conference on Computer Vision and Pattern
  Recognition Workshops}, 2019, pp. 0--0.

\bibitem{tygar2011adversarial}
J.~Tygar, ``Adversarial machine learning,'' \emph{IEEE Internet Computing},
  vol.~15, no.~5, pp. 4--6, 2011.

\bibitem{vargas2020openstreetmap}
J.~E. Vargas-Munoz, S.~Srivastava, D.~Tuia, and A.~X. Falcao, ``Openstreetmap:
  Challenges and opportunities in machine learning and remote sensing,''
  \emph{IEEE Geoscience and Remote Sensing Magazine}, vol.~9, no.~1, pp.
  184--199, 2020.

\bibitem{vavra2017recent}
P.~V{\'a}vra, J.~Roman, P.~Zon{\v{c}}a, P.~Ihn{\'a}t, M.~N{\v{e}}mec, J.~Kumar,
  N.~Habib, and A.~El-Gendi, ``Recent development of augmented reality in
  surgery: a review,'' \emph{Journal of healthcare engineering}, vol. 2017,
  2017.

\bibitem{wang2021daedalus}
D.~Wang, C.~Li, S.~Wen, Q.-L. Han, S.~Nepal, X.~Zhang, and Y.~Xiang,
  ``Daedalus: Breaking nonmaximum suppression in object detection via
  adversarial examples,'' \emph{IEEE Transactions on Cybernetics}, 2021.

\bibitem{wang2021dual}
J.~Wang, A.~Liu, Z.~Yin, S.~Liu, S.~Tang, and X.~Liu, ``Dual attention
  suppression attack: Generate adversarial camouflage in physical world,'' in
  \emph{Proceedings of the IEEE/CVF Conference on Computer Vision and Pattern
  Recognition}, 2021, pp. 8565--8574.

\bibitem{wang2021can}
W.~Wang, Y.~Yao, X.~Liu, X.~Li, P.~Hao, and T.~Zhu, ``I can see the light:
  Attacks on autonomous vehicles using invisible lights,'' in \emph{Proceedings
  of the 2021 ACM SIGSAC Conference on Computer and Communications Security},
  2021, pp. 1930--1944.

\bibitem{wang2019security}
X.~Wang, J.~Li, X.~Kuang, Y.-a. Tan, and J.~Li, ``The security of machine
  learning in an adversarial setting: A survey,'' \emph{Journal of Parallel and
  Distributed Computing}, vol. 130, pp. 12--23, 2019.

\bibitem{wengrowski2017reading}
E.~Wengrowski, K.~J. Dana, M.~Gruteser, and N.~Mandayam, ``Reading between the
  pixels: Photographic steganography for camera display messaging,'' in
  \emph{2017 IEEE International Conference on Computational Photography
  (ICCP)}.\hskip 1em plus 0.5em minus 0.4em\relax IEEE, 2017, pp. 1--11.

\bibitem{wengrowski2016optimal}
E.~Wengrowski, W.~Yuan, K.~J. Dana, A.~Ashok, M.~Gruteser, and N.~Mandayam,
  ``Optimal radiometric calibration for camera-display communication,'' in
  \emph{2016 IEEE Winter Conference on Applications of Computer Vision
  (WACV)}.\hskip 1em plus 0.5em minus 0.4em\relax IEEE, 2016, pp. 1--10.

\bibitem{Wu2020Defending}
\BIBentryALTinterwordspacing
T.~Wu, L.~Tong, and Y.~Vorobeychik, ``Defending against physically realizable
  attacks on image classification,'' in \emph{International Conference on
  Learning Representations}, 2020. [Online]. Available:
  \url{https://openreview.net/forum?id=H1xscnEKDr}
\BIBentrySTDinterwordspacing

\bibitem{wu2020making}
Z.~Wu, S.-N. Lim, L.~S. Davis, and T.~Goldstein, ``Making an invisibility
  cloak: Real world adversarial attacks on object detectors,'' in
  \emph{European Conference on Computer Vision}.\hskip 1em plus 0.5em minus
  0.4em\relax Springer, 2020, pp. 1--17.

\bibitem{xie2017mitigating}
C.~Xie, J.~Wang, Z.~Zhang, Z.~Ren, and A.~Yuille, ``Mitigating adversarial
  effects through randomization,'' \emph{arXiv preprint arXiv:1711.01991},
  2017.

\bibitem{xu2017feature}
W.~Xu, D.~Evans, and Y.~Qi, ``Feature squeezing: Detecting adversarial examples
  in deep neural networks,'' \emph{arXiv preprint arXiv:1704.01155}, 2017.

\bibitem{yan2016can}
C.~Yan, W.~Xu, and J.~Liu, ``Can you trust autonomous vehicles: Contactless
  attacks against sensors of self-driving vehicle,'' \emph{DEF CON}, vol.~24,
  no.~8, p. 109, 2016.

\bibitem{zhang2018perceptual}
R.~Zhang, P.~Isola, A.~A. Efros, E.~Shechtman, and O.~Wang, ``The unreasonable
  effectiveness of deep features as a perceptual metric,'' in \emph{CVPR},
  2018.

\bibitem{zhao2019seeing}
Y.~Zhao, H.~Zhu, R.~Liang, Q.~Shen, S.~Zhang, and K.~Chen, ``Seeing isn't
  believing: Towards more robust adversarial attack against real world object
  detectors,'' in \emph{Proceedings of the 2019 ACM SIGSAC Conference on
  Computer and Communications Security}, 2019, pp. 1989--2004.

\bibitem{zhou2018invisible}
Z.~Zhou, D.~Tang, X.~Wang, W.~Han, X.~Liu, and K.~Zhang, ``Invisible mask:
  Practical attacks on face recognition with infrared,'' \emph{arXiv preprint
  arXiv:1803.04683}, 2018.

\bibitem{zolfi2021translucent}
A.~Zolfi, M.~Kravchik, Y.~Elovici, and A.~Shabtai, ``The translucent patch: A
  physical and universal attack on object detectors,'' in \emph{Proceedings of
  the IEEE/CVF Conference on Computer Vision and Pattern Recognition}, 2021,
  pp. 15\,232--15\,241.

\end{thebibliography}
%\nocite{*}
% \input{Sources/appendix}
\end{document}